# An equilibrium model for RFP plasmas in the presence of resonant tearing modes with an application to particle and heat transport

P. Zanca$^{(*)}$, F. Sattin, E. Martines

*Consorzio RFX, Associazione Euratom-ENEA sulla Fusione*
*Corso Stati Uniti 4, 35127 Padova, ITALY*

**Abstract**

The equilibrium of a finite-β cylindrical RFP plasma in the presence of saturated-amplitude tearing modes is investigated. The singularities arising from the perturbative analysis of the MHD force balance equation $\boldsymbol{J}\times\boldsymbol{B}=\nabla p$ are resolved through a proper regularization of the cylindrical symmetric component of the pressure and parallel current density profiles, by setting to zero their gradient at the modes rational surfaces. An equilibrium model, which satisfies the regularization rule at the various rational surfaces, is developed. The comparison with the experimental data from the Reversed Field eXperiment (RFX) gives encouraging results. The model provides an easy tool to extrapolate the magnetic data, ordinarily measured only at the plasma edge, inside the plasma. As an application, we present the computation of the particle and heat stochastic diffusion coefficient from the reconstructed magnetic field profile.

## 1. Introduction

The study of the equilibrium configuration in Reversed Field Pinches (RFPs) is a far more difficult subject than in Tokamaks. The reason is that, in the former devices, the sustainment of the configuration against resistive diffusion is provided by the plasma itself through the generation of magnetohydrodinamical (MHD) modes (dynamo modes) [1]. Moreover it has been demonstrated that the stationary configuration cannot be axisymmetric [2] (this fact is commonly called Cowling's theorem). The result is that the RFP equilibrium configuration has to be recovered from a self-consistent, essentially three-dimensional (3-D) MHD calculation. A magnetic configuration characterized by saturated-amplitude MHD modes is described by the 3-D equilibrium force balance equation, whose simplest form is

---

$^{(*)}$ E-mail: zanca@igi.pd.cnr.it

1.1) $\mathbf{J} \times \mathbf{B} = \nabla p$

together with the Ampere's law and the divergence law.

1.2) $\nabla \times \mathbf{B} = \mu_0 \mathbf{J}$ ; $\nabla \cdot \mathbf{B} = 0$

Note that the flow effects related to the convective inertial term and the viscous force have been neglected in (1.1): our system of equations (1.1,1.2) describes an equilibrium in which the electromagnetic force dominates any other term in the motion equation. This simplifying hypothesis can be justified by the fact that in RFPs plasma the modes for sufficiently high amplitude are always phase locked together. According to the analysis discussed in [3] the phase-locking configuration is almost entirely determined by the electromagnetic force and only very weakly affected by plasma inertia and viscosity.

Present days RFPs are endowed with rather large aspect ratios, hence toroidicity-induced effects are expected to be subdominant and we will neglect them here. Thus, we choose to work in a cylindrical geometry: the system is taken to be periodic in the z-direction with periodicity length $2\pi R$, and a simulated toroidal angle $\phi = z/R$ is defined, so the coordinates are $(r, \theta, \phi)$.

A fully 3D treatment of Eqns. (1.1,1.2) would necessarily imply a numerical approach. Nevertheless the adoption of a perturbative method can be instructive as well. This is suggested by the fact that the dominant part of the magnetic field is cylindrically symmetric (zeroth-order component), and any deviation from axisymmetry can be regarded as a perturbation. Therefore, any field $A$ (which stands for $B^i$, $J^i$, or $p$) can be written as:

1.3) $A(r, \theta, \phi) = A_0(r) + a_1(r, \theta, \phi) + a_2(r, \theta, \phi) + ...$ ; $\quad a_i(r, \theta, \phi) = \sum_{m,n} a_i^{m,n}(r) e^{i(m\theta - n\phi)}$

where the terms $a_i$ indicates the perturbative order. The Fourier harmonics, associated to the plasma modes, are characterized by a poloidal mode number $m$, and a toroidal mode number $n$ ($n \neq 0$). The most important instabilities thought to exist in an RFP are tearing modes [4, 5]. They are resonant modes, i.e. a radius $r_s^{m,n}$ for which the cylindrical safety factor $q(r) \equiv r/R \, B_0^\phi(r)/B_0^\theta(r)$ satisfies the condition $q = m/n$, exists inside the plasma. The (zeroth-order) flux-surface where the previous condition holds is named $(m, n)$ rational surface. Therefore in the expansion (1.3) we will take into consideration only resonant harmonics. Tearing modes develop owing to the small but finite resistivity of the plasma. The resistivity breaks the flux-freezing constraint in the Faraday-Ohm's law, and allows the tearing and reconnection of equilibrium magnetic flux surfaces, to produce helical magnetic islands centered on each rational surface.

It is known that for arbitrary zeroth-order fields the resonant harmonics $a_i^{m,n}$ exhibit singularities at the modes rational surfaces. Nevertheless, a proper regulation of zeroth order field avoids the occurrence of such infinities [6]. The possibility of applying the regularization relies on the large latitude for the choice of the zeroth-order profiles. Moreover, the zeroth order field of the expansion (1.3) cannot be merely an arbitrary initial profile (such the one considered in the stability analysis), but must contain the modification produced by the presence of the saturated modes. It is in fact known that magnetic islands determine a localized flattening of the pressure and parallel current density profiles at the rational surfaces [7]. It will be shown that the local flattening is just the zeroth-order regularization required. This can have an important global effect on the profiles. In fact in the presence of many resonant modes the combined effect of the regularizations at the different rational surfaces determines an overall flattening of the profiles: the more modes we have, the flatter the profile we get. Tokamak plasmas are almost totally driven by external means and it is likely that the influence of instabilities on equilibrium profile be small there. On the other hand, in RFPs plasmas the dynamo modes determine a relaxation (flattening) of the equilibrium profiles which drive the configuration near the Taylor's minimum energy state [8] characterized by a constant parallel current density (normalized to the magnetic field). Therefore the regularization procedure seems a suitable way to quantify the relaxation of the zeroth-order profiles due to these modes. There are mainly two experimental observations that support this hypothesis: a) in the Reversed Field eXperiment (RFX) device [9], several resonant instabilities are observed and the parallel current profile is very flat. This is not only indicated by the standard $\alpha$-$\Theta_0$ equilibrium model [1], commonly adopted for RFPs, but it is also confirmed and reinforced by the internal polarimetric measurements [10]. b) When some of the secondary $m=1$ modes disappear, either for a spontaneous process ($\alpha$-mode [11]) or for an external influence (Pulsed Poloidal Current Drive-PPCD [12]), a <u>steepening</u> of the parallel current profile is observed. The same behaviour is observed for the pressure, even if only the electronic component can be measured [13]. Another confirmation of these ideas comes from the recently published reconstruction of the MST equilibrium profiles [14].

The plan of the work is the following: in Section 2 we analyze the problem of the equilibrium in the presence of saturated resonant instabilities, showing how to overcome the singularities by the ad-hoc regularization. For simplicity we will discuss in detail only the first perturbative order, giving a comment for the higher order corrections. In Sections 3 we show how the regularization rule leads in a natural way to a numerical model for computing, under some reasonable assumptions, the zeroth-order profiles. The required informations are the global equilibrium parameters $\Theta$, $F$, $\beta_p$, commonly adopted for the description of the RFPs equilibria, and the number of observed instabilities. Although to a large extent completely general, the actual numerical implementation of the algorithm relies here and there on specific features of the machine. We have chosen to adapt it to the RFX experiment. Section 4 is devoted to the description of the numerical algorithm adopted for solving equilibrium equations. Section 5 gives examples of application of the algorithm to specific RFX plasmas: zeroth-order profiles derived by our model are computed and compared with the standard α-$\Theta_0$ prediction. We also give an estimate for the first-order perturbation profiles and amplitudes inside the plasma, starting from the measurement of their values at the edge. The analysis [15] indicates that the perturbation profiles derived from the first order equations are a good approximation of the overall perturbation profile. This allows the spatial reconstruction of the total field perturbation. This part has been presented elsewhere [16]. The knowledge of the full magnetic field is essential also for an understanding of correlated quantities, e.g., transport. Particle transport in RFPs is known to be anomalous; the Rechester-Rosembluth theory relates it to the diffusive motion of magnetic field lines. In sections 6, 7 we accept this assumption, and use the computed magnetic field profiles to estimate transport coefficients, to be compared with estimates from the experiment. These results are entirely new. Section 8 summarizes the whole work, points out some less-developed points as well as possible improvements. Appendix A provides a complementary justification of the regularization rule within the island formalism.

**2. Equilibrium equations**

*2.1 Zeroth-order fields*

The zeroth-order equations are

2.1) $\mathbf{J}_0 \times \mathbf{B}_0 = \nabla p_0 \quad \nabla \times \mathbf{B_0} = \mu_0 \mathbf{J}_0$

where $\mathbf{B}_0 = [0, B_{0\theta}(r), B_{0\phi}(r)]$; $\mathbf{J}_0 = [0, J_{0\theta}(r), J_{0\phi}(r)]$. From (2.1) one gets

2.2) $\quad \nabla \times \mathbf{B}_0 = \mu_0 \dfrac{\mathbf{J}_0 \cdot \mathbf{B}_0}{B_0^2} \mathbf{B}_0 - \mu_0 \dfrac{\nabla p_0 \times \mathbf{B}_0}{B_0^2}$

Defining the normalized pressure gradient and the (normalized) parallel current density profile as

2.3) $\quad g(r) = \dfrac{\mu_0}{B_0^2}\dfrac{dp_0}{dr}$

2.4) $\quad \sigma(r) = \mu_0 \dfrac{\mathbf{J}_0 \cdot \mathbf{B}_0}{B_0^2}$

equation (2.2) provides the system

2.5) $\quad \dfrac{dB_{0\phi}}{dr} = -\sigma B_{0\theta} - g B_{0\phi}$

2.6) $\quad \dfrac{1}{r}\dfrac{d}{dr}(rB_{0\theta}) = \sigma B_{0\phi} - g B_{0\theta}$

In a real experiment it is unlikely that the pressure gradient could have values significantly greater than zero. Therefore we set $g(r) \leq 0$, so our pressure profiles are monotonic decreasing. Two important combinations of the equilibrium fields are

2.7) $\quad F^{m,n} = mB_{0\theta} - n\varepsilon B_{0\phi}$ ; $\quad \varepsilon(r)=r/R$

2.8) $\quad G^{m,n} = mB_{0\phi} + n\varepsilon B_{0\theta}$

### 2.2 First-order perturbations

The first order equations are

2.9) $\quad \mathbf{J}_0 \times \mathbf{b}_1 + \mathbf{j}_1 \times \mathbf{B}_0 = \nabla p_1$ ; $\quad \nabla \times \mathbf{b}_1 = \mu_0 \mathbf{j}_1$ ; $\quad \nabla \cdot \mathbf{b}_1 = 0$

Non-linear coupling of the perturbations appear in the higher order corrections. It is convenient taking the curl of the force-balance equation to get rid of the pressure term:

2.10) $\quad (\mathbf{b}_1 \cdot \nabla)\mathbf{J}_0 - (\mathbf{J}_0 \cdot \nabla)\mathbf{b}_1 + (\mathbf{B}_0 \cdot \nabla)\mathbf{j}_1 - (\mathbf{j}_1 \cdot \nabla)\mathbf{B}_0 = 0$

Equations (2.9, 2.10) allow for a straightforward Fourier analysis. By defining

2.11) $\psi^{m,n} = -irb_{r1}^{m,n}$

from the previous equations one gets the harmonics of the poloidal and toroidal field perturbations

2.12) $b_{\theta 1}^{m,n} = \dfrac{1}{m^2 + n^2\varepsilon^2}\left[-m\dfrac{d\psi^{m,n}}{dr} + n\varepsilon\sigma\psi^{m,n} + n\varepsilon g\dfrac{G^{m,n}}{F^{m,n}}\psi^{m,n}\right]$

2.13) $b_{\phi 1}^{m,n} = \dfrac{1}{m^2 + n^2\varepsilon^2}\left[n\varepsilon\dfrac{d\psi^{m,n}}{dr} + m\sigma\psi^{m,n} + mg\dfrac{G^{m,n}}{F^{m,n}}\psi^{m,n}\right]$

and the equation which gives the radial profile of $\psi^{m,n}$:

2.14) $F^{m,n}\left\{\dfrac{d}{dr}\left[\dfrac{r}{m^2+n^2\varepsilon^2}\dfrac{d}{dr}\psi^{m,n}\right] - \psi^{m,n}\left[\dfrac{1}{r} - \dfrac{r\sigma^2}{m^2+n^2\varepsilon^2} + \dfrac{2n\varepsilon(m\sigma+n\varepsilon g)}{(m^2+n^2\varepsilon^2)^2}\right]\right\} =$

$\psi^{m,n}\left\{\dfrac{1}{m^2+n^2\varepsilon^2}\left[rG^{m,n}\dfrac{d\sigma}{dr} - r\dfrac{d(gF^{m,n})}{dr} + \dfrac{2mn\varepsilon G^{m,n}g}{m^2+n^2\varepsilon^2} - rG^{m,n}\sigma g\right] + \dfrac{2n^2\varepsilon^2 g G^{m,n\,2}}{(m^2+n^2\varepsilon^2)^2 F^{m,n}}\right\}$

Moreover, writing the relation $\mathbf{B}\cdot\nabla p = 0$ for the *(m, n)* first-order harmonic we have:

2.15) $p_1^{m,n} = -\dfrac{B_0^2 g(r)}{\mu_0 F^{m,n}}\psi^{m,n}$

All the perturbations are therefore expressed in terms of $\psi^{m,n}$. By defining the quantity

2.16) $\chi^{m,n} = \left(\dfrac{r}{m^2+n^2\varepsilon^2}\right)^{1/2}\psi^{m,n}$

equation (2.14) is written in a more convenient form

$$2.17) \quad \frac{d^2\chi^{m,n}}{dr^2} - \chi^{m,n}\left[-\frac{m^4+10m^2n^2\varepsilon^2-3n^4\varepsilon^4}{4r^2(m^2+n^2\varepsilon^2)^2} + \frac{m^2+n^2\varepsilon^2}{r^2} - \sigma^2 + \frac{2n\varepsilon(m\sigma+n\varepsilon g)}{r(m^2+n^2\varepsilon^2)}\right] =$$

$$\frac{\chi^{m,n}}{F^{m,n}}\left\{G^{m,n}\frac{d\sigma}{dr} - \frac{d(gF^{m,n})}{dr} + \frac{2mn\varepsilon G^{m,n}g}{r(m^2+n^2\varepsilon^2)} - G^{m,n}\sigma g + \frac{2n^2\varepsilon^2 g}{r(m^2+n^2\varepsilon^2)}\frac{(G^{m,n})^2}{F^{m,n}}\right\}$$

This is the standard force-balance equation for a *(m,n)* radial field in finite-β cylindrical RFP. It is possible to demonstrate that this equation is equivalent to the one reported in [17].

*2.3 Resonant modes*

As written in the introduction, a mode *(m, n)* is resonant if the condition $F^{m,n}(r_s^{m,n}) = 0$ is satisfied for $0 < r_s^{m,n} < a$, where *a* is the plasma radius. Note that equations (2.12-2.17) hold for $F^{m,n} \neq 0$. Equation (2.17) must then be solved in the two separate regions between the axis and the rational surface $[0, r_s^{m,n}[$, and between the rational surface and the external boundary $]r_s^{m,n}, r_{shell}]$, which is assumed to be given by an ideal shell located at $r = r_{shell}$. Let us discuss the behaviour of the solution near $r_s^{m,n}$. Defining $x = r - r_s^{m,n}$, we take a Taylor expansion of $\sigma$ and *g* around *x* = 0:

$$2.18) \quad \sigma(x) = \sigma_0 + \sigma_1 x + o(x^2); \quad g(x) = g_0 + g_1 x + o(x^2)$$

For x → 0 equation (2.17) reduces to

$$2.19) \quad \frac{d^2}{dx^2}\chi = \chi\left[u + \frac{w}{x} + \frac{y}{x^2}\right]$$

where *u, w, y* are constant (for ease of notation we neglect the superscript *(m,n)*); in particular

$$2.20) \quad w = \frac{1}{(dF/dr)_{r_s}}\left[G\sigma_1 - g_0\frac{dF}{dr} + \frac{2mn\varepsilon G g_0}{r(m^2+n^2\varepsilon^2)} - G\sigma_0 g_0 + \frac{2n^2\varepsilon^2 G^2}{r(m^2+n^2\varepsilon^2)(dF/dr)}g_1 + \right.$$

$$\left. - \frac{2n^2\varepsilon^2 G^2(d^2F/dr^2)}{r(m^2+n^2\varepsilon^2)(dF/dr)^2}g_0\right]_{r_s}$$

2.21) $\quad y = \dfrac{1}{(dF/dr)^2} \left[ \dfrac{2n^2 \varepsilon^2 G^2}{r(m^2 + n^2 \varepsilon^2)} g_0 \right]_{r_s}$

The lowerscript $r_s$ means that all the quantities in the right-hand side must be evaluated at the location of the resonant surface. Also, note that $y \leq 0$ since $g_0 \leq 0$.

A regular solution of (2.19) is written as [18]:

2.22) $\quad \chi = |x|^\nu L(x)$

where

2.23) $\quad L(x) = L_0 + L_1 x \ln|x| + L_2 x + L_3 x^2 \ln|x| + o(x^2)$

and $\nu \geq 0$, $L_0 \neq 0$. Inserting in (2.19) one gets the following conditions

2.24) $\quad \nu(\nu - 1) = y$

2.25) $\quad \nu L_1 = 0$

2.26) $\quad L_1(1 + 2\nu) + 2\nu L_2 = L_0 w$

2.27) $\quad 2L_3(1 + 2\nu) = L_1 w$

Condition (2.24) provides two possible exponents, which are associated to the so-called "large" and "small" solution [18]:

2.28) $\quad \nu_L = \dfrac{1 - \sqrt{1 + 4y}}{2} \ ; \quad \nu_S = \dfrac{1 + \sqrt{1 + 4y}}{2} ; \quad \nu_L \leq \nu_S \ ; \quad 0 \leq \nu_{S,L} \leq 1$

To each of the two exponents is associated a solution, and therefore a set of coefficients $L_0, L_1, ...,$ which will be hereafter labeled with the further lowerscript "L" or "S". We suppose that the condition $1 + 4y > 0$ holds, otherwise the solution would exhibit oscillating singular behaviour for $x \to 0$. It is possible to show that this condition is just the Suydam criterion:

$$2.29) \quad \frac{rB_\phi^2}{8\mu_0}\left(\frac{q'}{q}\right)^2 > -p'$$

applied at the rational surface. This prescription makes the solution (2.22) (i.e. the radial field harmonic) a regular function approaching the rational surface.

We require, furthermore, that also the poloidal, toroidal field and pressure harmonics (2.12, 2.13, 2.15) be regular (i.e., finite) function for $x \to 0$. By inserting expressions (2.22, 2.23) into (2.12, 2.13, 2.15), we note two potentially singular terms; the former arises from the term including the derivative of $\chi$:

$$2.30) \quad \frac{d}{dx}\chi = \nu|x|^{\nu-1}\mathrm{sgn}(x)\cdot L(x) + |x|^\nu\left[L_1 \ln|x| + o(1)\right]$$

The second is due to the $1/F^{m,n}$ term: its contribution is of the form

$$2.31) \quad \frac{g_0}{x}\chi = g_0|x|^{\nu-1}\mathrm{sgn}(x)\cdot L(x)$$

Since these expressions are differently linear combined in (2.12, 2.13, 2.15) it is not possible that the divergences cancel out each other, and instead they must be regularized separately.

This implies $g_0 = y = 0$, otherwise $0 < \nu < 1$ (see (2.28)) and the terms with $|x|^{\nu-1}$ would be divergent. The condition $g_0 = y = 0$ imposes that the radial derivative of the zeroth-order pressure vanishes at the rational surface. Moreover, the request of having a monotonic pressure profile brings to the further condition $g_1 = 0$ (there must be a saddle-point, otherwise the rational surface would be the location for a local extremum).

Therefore the two possible exponents of (2.28) are

$$2.32) \quad \nu_L = 0, \quad \nu_S = 1.$$

The "large" solution gives a singular '$ln|x|$' term in (2.30). Therefore we set

$$2.33) \quad L_{1,L} = 0$$

which, together with (2.26, 2.27), yields

2.34) $w = 0; \quad L_{3,L} = 0$.

Therefore, the "large" solution reduces to

2.35) $\chi_L = L_{0,L} + L_{2,L} x + o(x^2)$ .

The condition $w = 0$, together with $g_0 = g_1 = 0$ implies $\sigma_1 = 0$: that is, the radial derivative of the parallel current profile also must vanish at the rational surface.

Note that the "large" solution, which has a finite radial field at the rational surface, is not compatible with the ideal form of the Faraday-Ohm's equation, for which $b_r \propto x \, \xi_r$ near the rational surface [19] ($\xi_r$ is the radial plasma displacement). Our discussion assumes therefore implicitly the effect of resistivity in the Faraday-Ohm's law.

Since we have imposed $w = 0$, for the "small" solution, using again Eqns. (2.25 - 2.27), one gets

2.36) $L_{1,s} = L_{2,s} = L_{3,s} = 0$

which gives

2.37) $\chi_s = x \left( L_{0,s} + o(x^2) \right)$

In conclusion the request that the first-order harmonics be finite when approaching the corresponding rational surface impose the following regularization conditions on the zeroth-order profiles

2.38) $\left. \dfrac{d\sigma}{dr} \right|_{r_s^{m,n}} = 0; \quad \left. \dfrac{dp_0}{dr} \right|_{r_s^{m,n}} = \left. \dfrac{d^2 p_0}{dr^2} \right|_{r_s^{m,n}} = 0;$

Equation (2.17) is solved imposing the suitable conditions at the origin and at the external boundary, and the natural requirement of the continuity of the solution $\psi$ at $r_s$. In general it will not be possible to solve the problem with a first radial derivative $d\psi/dr$ continuous at $r_s$. The jump of $d\psi/dr$ across the rational surface corresponds to the presence of a localized helical current sheet flowing there. (Note that such discontinuity imply a $\delta(r - r_S^{m,n})$ term in $d^2\psi/dr^2$, which is indeed allowed by equation (2.14) because $F^{m,n} \delta(r - r_S^{m,n}) = 0$ in the distribution sense)

*2.2 Higher-order perturbations*

The solution of the second-order equations

$$(2.39)\quad \mathbf{J}_0 \times \mathbf{b}_2 + \mathbf{j}_2 \times \mathbf{B}_0 + \mathbf{j}_1 \times \mathbf{b}_1 = \nabla p_2; \quad \nabla \times \mathbf{b}_2 = \mu_0 \mathbf{j}_2; \quad \nabla \cdot \mathbf{b}_2 = 0$$

is a much more complex task, due to the presence of non-linear coupling terms in the first-order perturbations. This has been carried out in ref. [20] for a zero-pressure equilibrium. The generalization to the finite-β case appears wearisome. Anyway, we are confident that the regularization previously described holds also for these higher order equations. In fact in the zero-pressure case equation (2.14) is modified just to incorporate the coupling terms (see eqns. (A43-A61) of [20]). These extra terms are potentially singular at the modes rational surfaces, but they are all proportional to $d\sigma/dr$. Therefore these extra singularities can be healed by the same prescription here described, i.e. flattening $\sigma(r)$ near the rational surfaces.

### 3. A model for current and pressure profiles

According to the previous discussion, in the presence of saturated resonant modes the quantities $d\sigma/dr$, $g(r)$ and $dg/dr$ must vanish at the mode rational surfaces. In RFX we have $m=0$ modes resonant at the $q = 0$ surface (the reversal surface), and many $m=1$ modes whose rational surface is inside the reversal surface (internally resonant modes). Therefore we make the *ansatz*

$$(3.1)\quad \frac{d\sigma}{dr}(r) = M(r)f(r), \quad g(r) = M^2(r)h(r)$$

where $M(r)$ is the "regularization" term

$$(3.2)\quad M(r) = q(r) \prod_{n_{\min}}^{n_{\max}} (1 - nq(r))$$

which automatically satisfies $d\sigma/dr = g(r) = dg/dr = 0$ at the $m=1$ and $m=0$ mode rational surfaces, provided that the shape functions $f(r)$, $h(r)$ (till now undefined) are there regular functions. This regularization prescription is similar to that discussed in [6]. If we want a monotonic decreasing $\sigma(r)$ profile then we set

$$(3.3) \quad M(r) = \left| q(r) \prod_{n_{min}}^{n_{max}} (1 - nq(r)) \right|$$

together with the supplementary constrain $f(r) \leq 0$. Note that, due to the modulus, in this case the second derivative of $\sigma$ is not defined at the rational surfaces. Nevertheless the expressions (2.18-2.20) require the first derivative only. At $r = 0$, for symmetry reasons, we have $d\sigma(0)/dr = 0$ and $g(0) = 0$. This implies $f(0) = h(0) = 0$.

Note that $M$ is indeed a function of $q$; as long as $q(r)$ is a monotonic function, we can use $q$ in place of $r$ as independent variable and write

$$3.4) \quad f(r) = \frac{dq}{dr} w(q), \quad h(r) = \frac{1}{B_0^2} \frac{dq}{dr} u(q)$$

($w$ and $u$ must not be confused with the same symbols used in Sec. 2: there, they were numbers, here they stand for functions of the radial coordinate). Being $dq(0)/dr = 0$, the symmetry conditions at $r = 0$ are automatically satisfied. From (3.1, 3.4) we get

$$3.5) \quad \frac{d\sigma}{dq} = M(q)w(q) \rightarrow \sigma(q) - \sigma(q_a) = \int_{q_a}^{q} d\overline{q} M(\overline{q}) w(\overline{q})$$

$$3.6) \quad \mu_0 \frac{dp_0}{dq} = M^2(q)u(q) \rightarrow \mu_0 p_0(q) - \mu_0 p_0(q_a) = \int_{q_a}^{q} d\overline{q} M^2(\overline{q}) u(\overline{q})$$

where $q_a$ is the value at the plasma boundary, in RFX determined by a graphite wall placed at the radius $a = 0.457$ m. The graphite wall leans against an inconel vessel located at $r = r_V$, beyond which there is a vacuum region which extends up to the conducting shell placed at $r_{shell} = 1.17 \cdot a$. In a situation of stationary modes (which is the standard situation in RFX) helical eddy currents cannot be induced in the vessel, so both $\psi$ and $d\psi/dr$ are continuous at $r = r_V$. There is still the possibility for zeroth-order current to flow in the inconel vessel; our fluid model fails there, but we are confident that its amplitude is so small that it can be safely neglected. Furthermore zeroth-order currents cannot flow at all in the graphite wall [21]. In conclusion we treat the entire region between the plasma boundary and the shell as vacuum:

(3.7)    $\sigma(r) = 0$, $g(r) = 0$  for  $a < r < r_{shell}$

Notice that if we want equation (2.14) to hold at the boundary $r = a$, then we need the quantities $\sigma$, $d\sigma/dr$, $g(r)$, $dg/dr$ to be defined there. The condition (3.7) then implies

(3.8)    $\sigma(a) = g(a) = d\sigma(a)/dr = dg(a)/dr = 0$

From $g(a) = 0 = dp_0(a)/dr$ and $p_0(r>a) = 0$ it follows $p_0(q_a) = 0$. Moreover, conditions (3.8) force the following constraints on the functions $w$, $u$:

(3.9)  $w(q_a) = 0; \quad u(q_a) = 0 = \left.\dfrac{du}{dq}\right|_{q_a}$

The model depends upon some parameters to be chosen on the basis of supplementary assumptions: in RFX the zeroth-order profiles are characterized by the three experimental parameters

(3.10)  $\Theta = \pi a^2 \dfrac{B_{0\theta}(a)}{\Phi_t(a)}; \quad F = \pi a^2 \dfrac{B_{0\phi}(a)}{\Phi_t(a)}; \quad \beta_p = \dfrac{4\mu_0 \int_0^a p_0(r)r\,dr}{a^2 B_{0\theta}^2(a)}$

where $\Phi_t$ is the toroidal flux. All the edge magnetic quantities are provided by a direct measure, so we can consider the determination of $\Theta$, $F$ to be exact. Instead, the integral which appears in the expression for $\beta_p$ depends on the assumption made on the pressure profile. At present, in RFX $\beta_p$ is computed assuming given polynomial expressions for the electron density and temperature profiles. Our pressure model is different, so we take this "experimental" $\beta_p$ only as a reference value. Note that

(3.11)  $q_a = \dfrac{a}{R}\dfrac{F}{\Theta}$

so $q_a$ can be used in place of F or $\Theta$.

If we require that different triplets *(F, Θ, β*$_p$*)* correspond to different profiles, the number of free parameters must be at least three. Another element to take into consideration is the Suydam criterion (2.29) for the pressure gradient. In our model this criterion is verified in most of the plasma, due to the prescriptions (3.1-3.2) which flatten the pressure profile on a wide region. A violation could arise in the edge zone, because in our model the pressure gradient is mostly concentrated there. Anyway we have no elements to say that in RFX the Suydam criterion is everywhere fulfilled, and we cannot indeed exclude the presence of localized interchange modes at the very edge of the plasma. This discussion suggests that a fourth free parameter is needed to model the pressure profile near the plasma boundary: this parameter will be tuned in order satisfy, at least marginally, the Suydam criterion there.

According to these arguments and to the condition (3.9), the simplest model for the functions *w, u* yields the expressions

$$(3.12) \quad w(q) = w_0 \left(1 - \frac{q}{q_a}\right)^{\xi} ; \quad u(q) = u_0 \left(1 - \frac{q}{q_a}\right)^{\eta} ; \quad \xi > 0 ; \quad \eta > 1;$$

where the four free parameters are $q_a$, the normalization constant $u_0$ and the exponents $\xi, \eta$. The normalization constant $w_0$ is not free but depends on $(q_a, \xi)$ as well as on the definition of the regularization term *M(r)*. This will be demonstrated in the following discussion.

### 4. Getting the solution: numerical scheme

*4.1 Computation of the zeroth-order profiles*

The actual procedure to obtain the zeroth-order profiles for a given triplet *(F, Θ, β*$_p$*)* or *($q_a$, F, β*$_p$*)* is quite involved and in principle, there is not guarantee for the solution to be unique, though we found that all of the admissible solutions are very close between them.

*4.1.1 Definition of M(r)*

First of all, the two possible choices (3.2, 3.3) for the regularization term *M(r)* discriminate between monotonic and not-monotonic *σ(r)* profiles. An integrated analysis [10] of the external magnetic signals and the data provided by a five-chord infrared (FIR) polarimeter indicates that in RFX the *σ(r)* profiles should be <u>very flat</u>, or even <u>hollow</u> with a maximum in the external region of the plasma. In our model the flattening of the *σ(r)* is a direct consequence of the regularization term *M(r)*. Moreover with the non-monotonic choice (3.2) we have a local maximum of *σ(r)* just at the reversal surface where *q(r)=0*. We will investigate both possibilities (3.2, 3.3).

In actual calculations, the product over *(1, n)* modes must -of course- be truncated to a finite number of terms. The number of *m=1* factors in *M(r)* is determined by the experimentally observed modes. Generally in RFX the dominant internally resonant *m=1* modes have a toroidal number in the range *n=7÷12*. There is also a tail of secondary modes which in some pulses can also extend to high (*n≈18*) mode numbers. Taking into account that the flattening of the computed $\sigma(r)$, $p_0(r)$ profiles extends beyond the resonance position of the *m=1*, *n= $n_{max}$* mode (this effect is more pronounced for the pressure, because we have a term $M^2(r)$ there), the range [$n_{min}$, $n_{max}$] of factors in *M(r)* is chosen in order to include <u>all</u> of the dominant modes and <u>most</u> of the secondary modes, with the check *a posteriori* that the flat region of $\sigma(r)$ and $p_0(r)$ includes all of the rational surfaces of the modes experimentally observed. The latter condition requires a lower number of factors for *g(r)* than for $\sigma(r)$, so we leave the possibility that $n_{max}$ appearing in the definition of *M(r)* could be different for $\sigma(r)$ and *g(r)*. In conclusion, there is some freedom in the definition of *M(r)*, but different plausible choices of the *m=1* factors do not substantially change the profiles.

### 4.1.2 Computation of q(r) and σ(r)

Having defined *M(r)*, we set $q_a$ equal to the experimental value (3.11), and guess a value $\xi>0$ for the exponent of *w(q)*. First of all we compute the *q(r)* and $\sigma(r)$ profiles: in fact the two equations (2.5-2.6) can be combined to give a single equation for *q(r)* where the pressure gradient term *g(r)* does not appear. We combine this equation with the expression for $d\sigma/dr$ to form the system

4.1) $\begin{cases} \dfrac{dq}{dr} = \dfrac{2}{r} q \left[ 1 - \dfrac{R\sigma}{2} q \right] - \dfrac{r}{R} \sigma \\ \dfrac{d\sigma}{dr} = w_0 M(r) \dfrac{dq}{dr} \left( 1 - \dfrac{q}{q_a} \right)^{\xi} \end{cases}$

The system is solved in the interval *[0, a]*. The apparently more natural direction of integration is from the edge towards the centre, since one knows the boundary conditions $q(a) = q_a$, $\sigma(a) = 0$. Note that the condition $\sigma(a) = 0$ forces the normalization constant $w_0$ to be (see (3.5, 3.12)):

4.2) $w_0 = \dfrac{\sigma(0)}{\int_{q_a}^{q_0} d\bar{q} M(\bar{q})(1 - \bar{q}/q_a)^{\xi}}$

that is, the r.h.s. of the latter of (4.1) depends upon the unknown value of $\sigma$ at the centre. For this reason it is more convenient to solve eqns. (4.1) in the direction from the centre towards the edge, using guessed initial values $q_0$, $\sigma(0)$, and iterating until the matching at the edge with the boundary conditions is obtained. To this purpose, remember that the relation

$$4.3) \quad q_0 = \frac{2}{R\sigma(0)}$$

must be satisfied. Of course, the value of $\sigma(0)$ must also be consistent with the resonant modes taken in the product (3.2), so the condition $2(n_{min}-1)/R < \sigma(0) < 2n_{min}/R$ must hold. We just mention further that the r.h.s of the former of (4.1), in the limit $r \to 0$, takes the form $0/0$. Thus, some care must be exerted in handling this limit.

### 4.1.3 Computation of $B_{0\phi}(r)$, $B_{0\theta}(r)$

The $\sigma(r)$ and $q(r)$ profiles obtained from (4.1) are then used to determine the zeroth-order magnetic fields by solving equations (2.5, 2.6). The pressure gradient $g(r)$ is given by

$$4.4) \quad g(r) = u_0 M^2(r) \frac{1}{B_0^2(r)} \frac{dq(r)}{dr} \left(1 - \frac{q(r)}{q_a}\right)^{\eta}.$$

Remember that the pressure profile is

$$4.5) \quad \mu_0 p_0(r) = u_0 \int_{q_a}^{q(r)} d\bar{q} M^2(\bar{q}) \left(1 - \frac{\bar{q}}{q_a}\right)^{\eta}.$$

In (4.4) and (4.5) $q(r)$ is just the solution derived from (4.1). The parameter $u_0$ is tuned in order to have from the last of (3.10) a $\beta_p$ close to the experimental estimated value. The exponent $\eta$ has little influence on $\beta_p$. It only determines the pressure gradient at the plasma edge and is chosen by comparison with the Suydam criterion.

In general the F and $\Theta$ parameters computed from the solutions $B_{0\phi}(r)$, $B_{0\theta}(r)$ are close (they stand in the correct ratio) but not equal to the experimental values. To match the desired values it is sufficient to change slightly the exponent $\xi$ or the number of m=1 resonance ( i.e. $n_{max}$ ) and to repeat the procedure starting from equation (4.1).

To summarize, our procedure involves solving equations (2.5, 2.6, 4.1) through fitting some free parameters to experimental quantities. Roughly speaking, we can sketch the following recipe:

| **Constraints** | **Related parameter** |
|---|---|
| Zero gradient at the mode resonances | Regularization term $M(r)$: $n_{min}$, $n_{max}$ |
| $F$, $\Theta$ | $q_a$, $\xi$ |
| $\beta_p$. | $u_0$ |
| Suydam criterion | $\eta$ |
| Boundary condition: $\sigma(a)=0$. | $w_0$ |

Table 1. Left column: constraints to be satisfied by the model functions. Right column: terms which are mostly related to the corresponding request of the left column.

We remark that Table I must be considered just as a quick reminder: actually, varying any of the quantity in the right-hand column affects several quantities in left-hand column.

It is convenient to adopt the following normalization for the zeroth-order quantities when working with the above equations (we remind again RFX parameters: $R=2$ m, $a=0.457$ m):

4.6) $\quad \hat{r} = r/a; \quad \hat{R} = R/a; \quad \hat{g} = g \cdot a; \quad \hat{\sigma} = \sigma \cdot a;$

4.7) $\quad \hat{B}_\phi = B_{0\phi}/B_{0\phi}(0); \quad \hat{B}_\theta = B_{0\theta}/B_{0\phi}(0);$

4.8) $\quad \hat{F}^{m,n} = F^{m,n}/B_{0\phi}(0); \quad \hat{G}^{m,n} = G^{m,n}/B_{0\phi}(0);$

The normalized pressure and the Suydam criterion are written as

4.9) $\quad \hat{p} = \dfrac{\mu_0 p_0(r)}{B_{0\phi}^2(0)}; \quad -\dfrac{d\hat{p}}{d\hat{r}} < \dfrac{\hat{r}\hat{B}_\theta^2}{8\varepsilon^2}\left(\dfrac{dq}{d\hat{r}}\right)^2;$

From now on, any tilded variable ($\hat{X}$) will implicitly refer to a normalized quantity.

### *4.2 First-order perturbation quantities*

As remarked in the introduction the first-order perturbation can be taken as a good approximation of the 'true' perturbation [15]. The radial profile of the various *(m, n)* first-order harmonics is obtained from eq. (2.17), which is solved using the zeroth-order profiles for $\sigma(r)$ and $g(r)$, computed according to the method just outlined. The edge boundary conditions, which determine the perturbation amplitudes, are given by the experimental measurements, in RFX available at the shell:

$$4.10) \quad \chi^{m,n}(\hat{r}_{shell}) = 0; \quad \frac{d\chi^{m,n}(\hat{r}_{shell})}{d\hat{r}} = \frac{\hat{r}_{shell}^{1/2}\left(m^2 + n^2 \varepsilon_{shell}^2\right)^{1/2}}{n\varepsilon_{shell}} a^{3/2} b_\phi^{m,n}(\hat{r}_{shell});$$

The first condition (4.10) is strictly true only for an ideal shell. The RFX shell is thick but not exactly ideal, since a slow penetration of the radial field is observed. This is shown by the measurements of two poloidal arrays of radial field pick-up coils placed on the inner surface of the shell. At present toroidal arrays of radial field probes, which are necessary to obtain the harmonics $b_r^{m,n}$ (proportional to $\chi^{m,n}$) at a given radius, are not available. Therefore the simplest way to set the initial value for $\chi^{m,n}$ is to adopt the ideal shell approximation (assumed in all of the following examples).

The second condition (4.10) is derived by a combination of (2.13, 2.16) applied in vacuum, and by the ideal shell constraint $\chi^{m,n}(r_{shell})=0$. The harmonic $b_\phi^{m,n}$ is indeed <u>measured</u> at the shell (for the $m=0, 1,$ modes) by two toroidal arrays of 72 equally spaced toroidal field pick-up coils placed at opposite poloidal angles [22].

The condition near the origin is $\chi^{m,n}(\hat{r}) \propto \hat{r}^{|m|+1/2}$ for $m \neq 0$ and $\chi^{0,n}(\hat{r}) \propto \hat{r}^{3/2}$ for $m=0$. The phase $\varphi^{m,n}$ of the harmonic $\psi^{m,n}$ is taken constant with respect to $r$, so we can write

$$4.11) \quad \psi^{m,n}(r,t) = \tilde{\psi}^{m,n}(r,t) e^{i\varphi^{m,n}(t)}.$$

Note that the mode amplitude $\tilde{\psi}^{m,n}$ is a real solution of equation (2.14). By integrating Ampere's law

$$4.12) \quad \mu_0 j_\theta^{m,n} = -i\frac{n}{R} b_r^{m,n} - \frac{\partial b_\phi^{m,n}}{\partial r}; \quad \mu_0 j_\phi^{m,n} = \frac{1}{r}\frac{\partial (r b_\theta^{m,n})}{\partial r} - i\frac{m}{r} b_r^{m,n};$$

in a narrow region $[r_s^{m,n} -\delta, r_s^{m,n} +\delta]$ around the mode rational surface, we can get the poloidal and toroidal components of the helical current sheet which flows there:

4.13) $\mu_0 \delta J_{\theta 1}^{m,n} \equiv \mu_0 \int_{r_s^{m,n}-\delta}^{r_s^{m,n}+\delta} j_{\theta 1}^{m,n} dr = -b_{\phi 1}^{m,n}\Big|_{r_s^{m,n}-\delta}^{r_s^{m,n}+\delta} + o(\delta) = -\frac{n\varepsilon}{(m^2+n^2\varepsilon^2)}\frac{d\tilde{\psi}^{m,n}}{dr}\Big|_{\hat{r}_s^{m,n}-}^{\hat{r}_s^{m,n}+} + o(\delta)$

4.14) $\delta J_{\phi 1}^{m,n} \cong \frac{m}{n\varepsilon} \delta J_{\theta 1}^{m,n}$

In the last equality of (4.13) we have taken into account that the third term in the square bracket of (2.12, 2.13) is zero at the rational surface, because $g(r)$ has a second order zero there.

### 5. Examples of application of the method
#### 5.1 Zeroth-order profiles: monotonic σ(r)

Table 2 reports some examples of zeroth-order profiles computations with the assumption of a monotonic $\sigma(r)$, therefore using the definition (3.3) for $M(r)$. Note that in this case $w_0 > 0$, so $d\sigma(r)/dr$ is negative everywhere, being $dq/dr<0$.

| Shot | F | Θ | βp | (1, n)σ | (1, n) g | $\hat{\sigma}(0)$ | ξ | $\hat{u}_0$ | η |
|---|---|---|---|---|---|---|---|---|---|
| 11029 t=30 | -0.119 | 1.408 | 0.060 | 7-14 | 7-12 | 2.83 | 0.4 | 2200 | 1.3 |
| 11029 t=35 **PPCD** | -0.224 | 1.476 | 0.056 | 7-10 | 7-9 | 2.844 | 0.05 | 330 | 1.2 |
| 8763 t=66 | -0.163 | 1.416 | 0.060 | 7-13 | 7-13 | 2.7876 | 0.35 | 1100 | 1.8 |
| 8763 t=73 **α-mode** | -0.217 | 1.472 | 0.052 | 7-10 | 7-10 | 2.854 | 0.06 | 450 | 1.6 |
| 8073 t=31.5 | -0.251 | 1.49 | 0.090 | 7-14 | 7-14 | 2.792 | 0.6 | 250 | 2.2 |

Table 2. Plasma parameters as guessed from our model for a set of experimental conditions. F, Θ, and poloidal beta are from experiment and must be matched by the model. The other columns contain, from left to right: the modes retained in the computation of σ and of pressure, the fitting parameters appearing in Eqns (4.1-4.4).

The fifth and sixth columns contain the range of $m=1$ factors used in the $M(r)$ terms for $\sigma(r)$ and $g(r)$ respectively. The $m=0$ factor is always present.

All the calculations have been performed with the "*Mathematica 4.0*" software. Note that the exponents $\xi, \eta$ do not vary too much from shot to shot. In the first column the time is given in *ms*.

Our model provides profiles which fit the global/edge measured parameters *(F, Θ, β_p)*. Comparing our results with whole radial shape of the profiles is at present impossible, since in RFX we have not a shot-by shot measurement of these profiles. Anyway, as told in sec 4.1, the first experimental reconstruction of *σ(r)* indicates a flat profile (see fig. 4 of Ref. [10]), which is largely compatible with the plots shown below. The same is true for the pressure, even if only the electronic component is measured (see fig. 4 of Ref. [13]).

Pulse 11029 at t=30ms

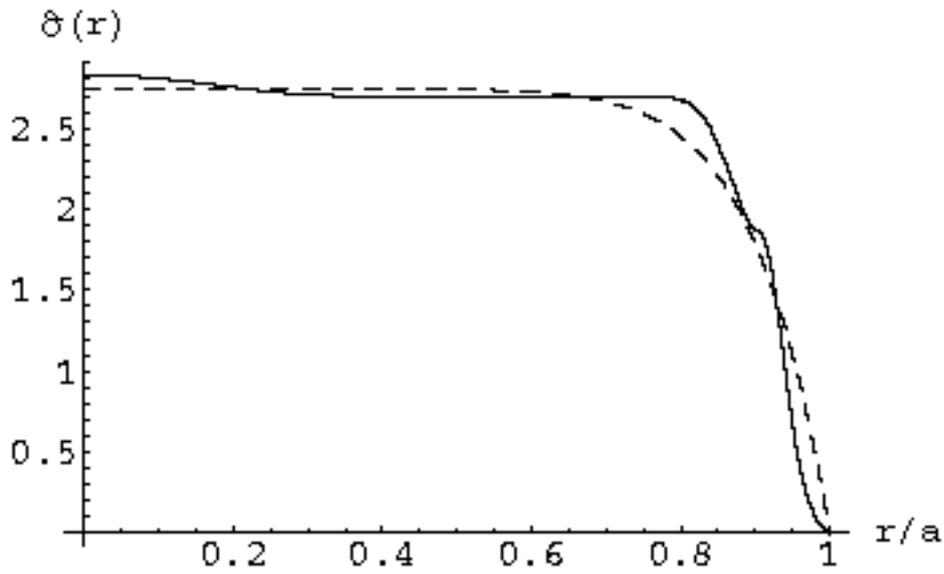

Fig. 1. The normalized σ profile. Pulse 11029 at t=30ms (standard conditions). Continuous line present model; dashed-line α-Θ$_0$ model

In Fig. 1 the continuous line is the result of our model while the dashed line is the reconstruction provided by standard α-Θ$_0$ model [1],

$$5.1) \quad \sigma(r) = \frac{2\Theta_0}{a}\left(1 - \left(\frac{r}{a}\right)^\alpha\right)$$

The differences between the two models arise in the central part of the plasma, where our model predicts a weak increase of *σ(r)* towards zero, and in the external region where we obtain a more extended flat zone, a further flattening in correspondence of the *m=0* resonance *(r≈ 0.9a)* and a smoother behaviour near the plasma boundary. The flat region *0.3 < r < 0.8* corresponds to the location of *m=1* modes. Figs. (2-4) show the other fundamental quantities calculated within the model.

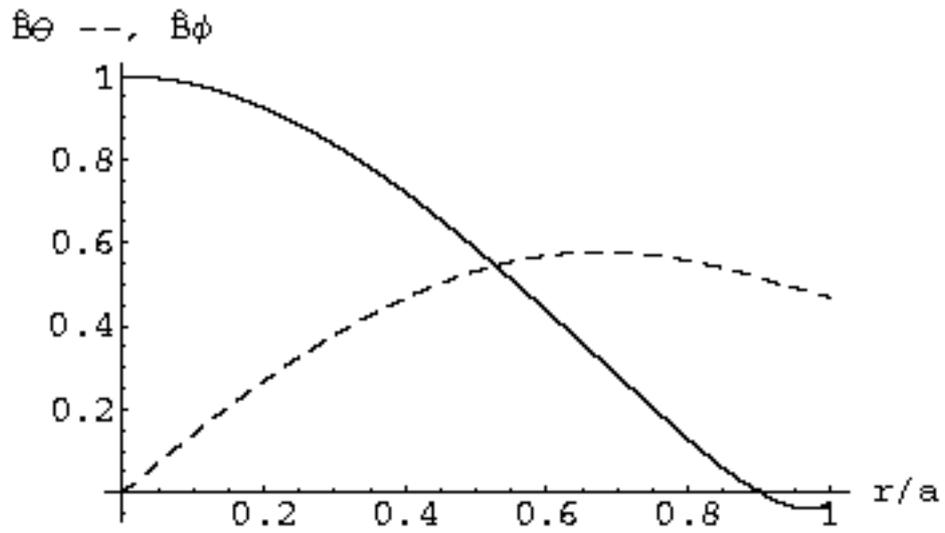

Fig. 2: The normalized toroidal and poloidal field profiles.

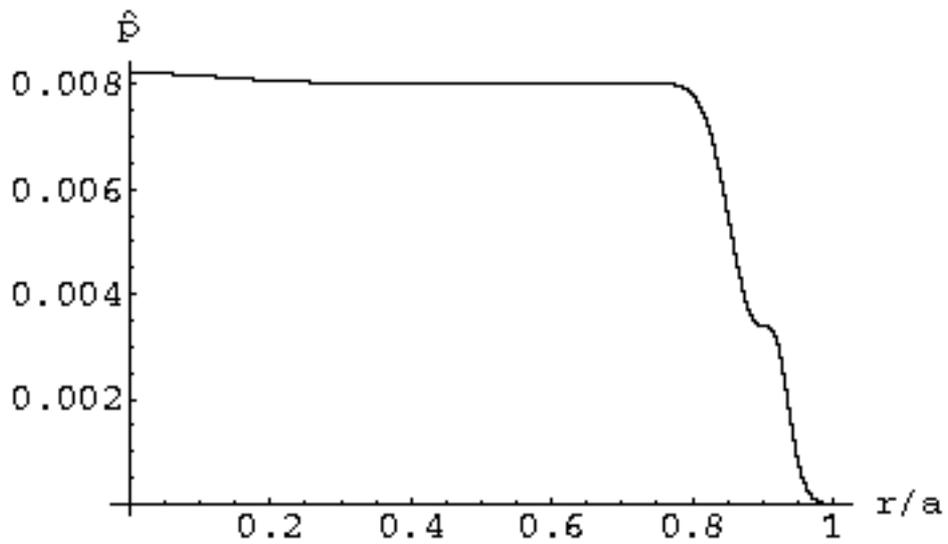

Fig. 3: The normalized pressure profile. The flattening due to the *m=0* modes at *r≈0.9a* is clearly evident

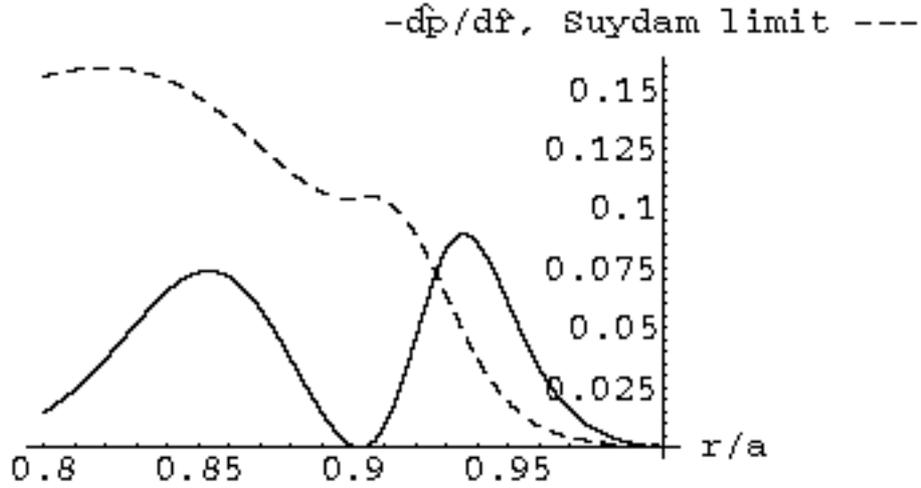

Fig. 4. Here the comparison, in the external region of the plasma, between the normalized pressure gradient (continuous line) and the Suydam limit (dashed line) is shown: the violation of this limit is at the very edge. The agreement could be improved by increasing the exponent η.

Pulse 11029, t = 35 ms, during PPCD

The same shot is considered when the PPCD [12] is active. During the PPCD some of the secondary *m=1* modes disappear. This entails a reduction of $n_{max}$ in *M(r)* and a consequent steepening of the σ*(r)* profile (fig. 5).

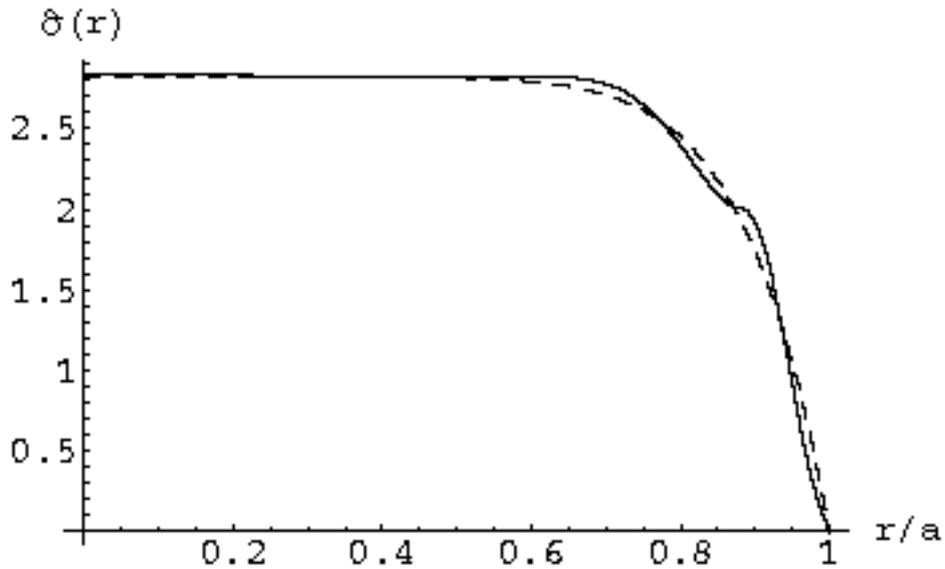

Fig. 5. The normalized σ profile. Pulse 11029 during PPCD. Continuous line present model; dashed-line α-$\Theta_0$ model

The peaking of the profile is also predicted by the α-$\Theta_0$ model. A similar peaking is apparent in the pressure profile, and it is also observed experimentally [13]. Moreover the pressure gradient at the edge is found to decrease in our model (fig.6). As discussed is the introduction this confirms the existence of a connection between the number of the instabilities and the shape of the zeroth-order profiles. During the PPCD the activity of the dynamo modes is weaker and consequently the profile is less relaxed (steeper).

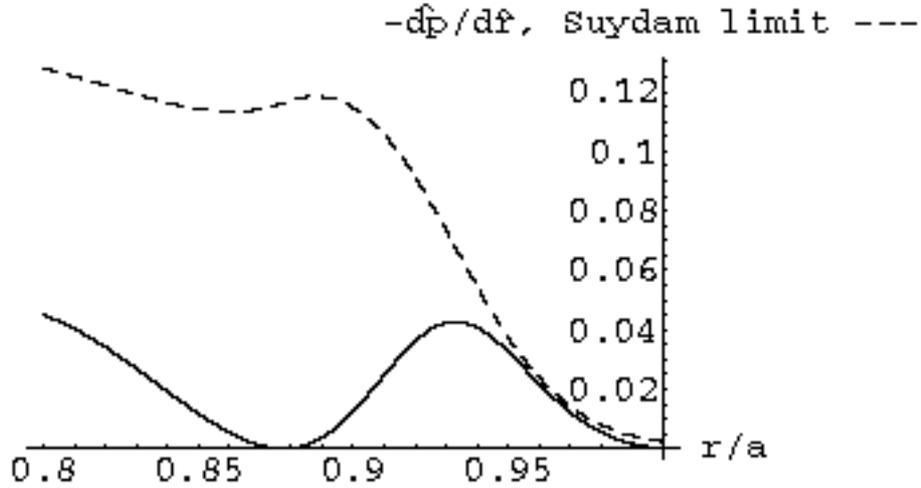

Fig. 6. During the PPCD our model satisfies the Suydam criterion also at the very edge of the plasma.

Alike the PPCD, the so called α-mode is characterized by a decrease of the secondary m=1 modes and by a steepening of the profiles. In the α-mode reported in the table our model satisfies the Suydam criterion also at the very edge of the plasma.

*5.2 Zeroth-order profiles: non-monotonic $\sigma(r)$*

As mentioned before in sec.4.1, there are evidences supporting the possibility for σ profiles to be non-monotonic in RFX plasmas. In our formalism, it means using Eq. (3.2) instead of (3.3) in computations. Note that in this case $w_0<0$, because of the significant negative contribution of *M(r)* to the integral (4.2) in the region between the reversal and the edge. With this kind of profile we found that the convergence of equations (2.5, 2.6, 4.1) to the experimental desired values of *(F, Θ, $β_p$)* is not guaranteed. We relate this to the following physical aspects of the problem:

(I) we have found that a necessary condition is a decreasing *σ(r)* between the origin and the first *m=1* rational surface. With the monotonic profile this condition is automatically fulfilled. Instead, using the definition (3.1) it is satisfied only with an odd number of *m= 1* mode. This suggest that the non-monotonic model for *σ(r)* should be refined including an extra factor, e.g.:

$$5.2) \quad \frac{d\sigma}{dr} = w_0 q(r) \prod_{n_{min}}^{n_{max}} (1 - nq(r)) \frac{dq}{dr} \left(1 - \frac{q}{q_a}\right)^{\xi} \cdot \left[\text{sgn}(1/n_{min} - q)\right]^{n_{max} - n_{min}}$$

(II) There should not be too much distance between the rational surface corresponding to $n_{max}$ and the reversal position, otherwise $\sigma(r)$, which has a positive derivative in this region, could grow to unreasonable values at the reversal. This requires a configuration with a considerable number of secondary $m=1$ modes, in other word a sufficiently relaxed profile.

(III) Since in the non-monotonic case the maximum of $\sigma(r)$ is reached at the reversal surface and it is unlikely that too much current be driven at the very edge of the plasma, the reversal should be at a suitable distance from the plasma boundary, i.e. the $F$ parameter should be sufficiently negative, to make the profile plausible.

Table 3 lists some specific cases. From fig.7 one can notice a significant departure from the $\alpha$-$\Theta_0$ reconstruction. Figure 8 instead compares the monotonic and non-monotonic models.

| Shot | F | $\Theta$ | $\beta p$ | $(1,n)\sigma$ | $(1,n)g$ | $\hat{\sigma}(0)$ | $\xi$ | $\hat{u}_0$ | $\eta$ |
|---|---|---|---|---|---|---|---|---|---|
| 8073 t=31.5 | -0.251 | 1.49 | 0.09 | 7-15 | 7-13 | 2.832 | 0.96 | 550 | 2.4 |
| 8069 t=40 | -0.391 | 1.58 | 0.08 | 7-13 | 7-11 | 2.875 | 1. | 120 | 2 |
| 14170 t=40 | -0.249 | 1.463 | 0.06 | 7-14 | 7-12 | 2.838 | 1. | 350 | 2 |

Table 3. Same symbols of Table 2 but for the non-monotonic current profile.

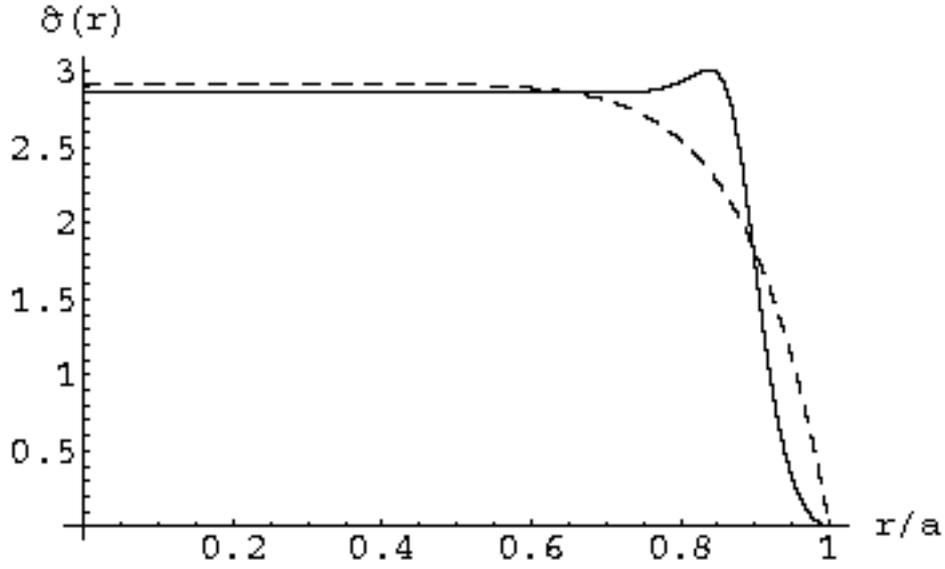

Fig. 7: Shot 8069, t=40ms. Continuous line present model; dashed-line α-Θ$_0$ model.

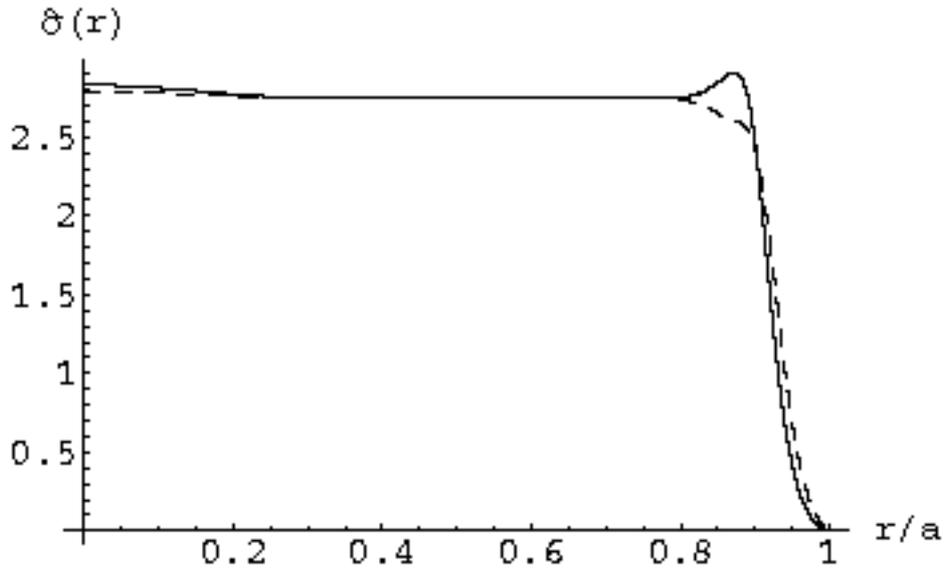

Fig. 8: Shot 8073, t=31.55ms The two kind of profiles admissible in our model, monotonic (dashed line) and not-monotonic (continuous line), are compared. In this case there is not a substantial difference apart in correspondence of the *m*=0 mode.

*5.3 Examples of perturbation profiles*

Let's examine the radial profile of the first-order perturbation associated to the typical dominant *m=1, n=8* mode. Figures 9, 10, 11, feature respectively the amplitudes $\tilde{\psi}^{1,8}(r)$, $\tilde{b}_{r1}^{1,8}(r)$, $\tilde{b}_{\phi 1}^{1,8}(r)$ in the pulse 11029 at t=30ms. The zeroth-order profile has already been determined (second row of Table 2). It is indeed a huge amplitude mode, with an edge measured value $b_\phi^{1,8}(1.17)$= -6.2mT.

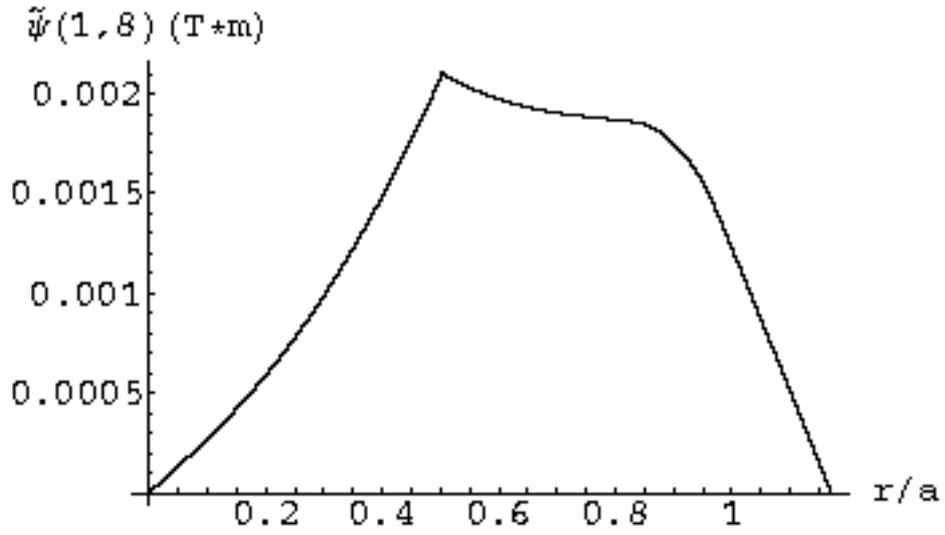

Fig. 9: Shot 11029 t=30ms. The units are (T·m)

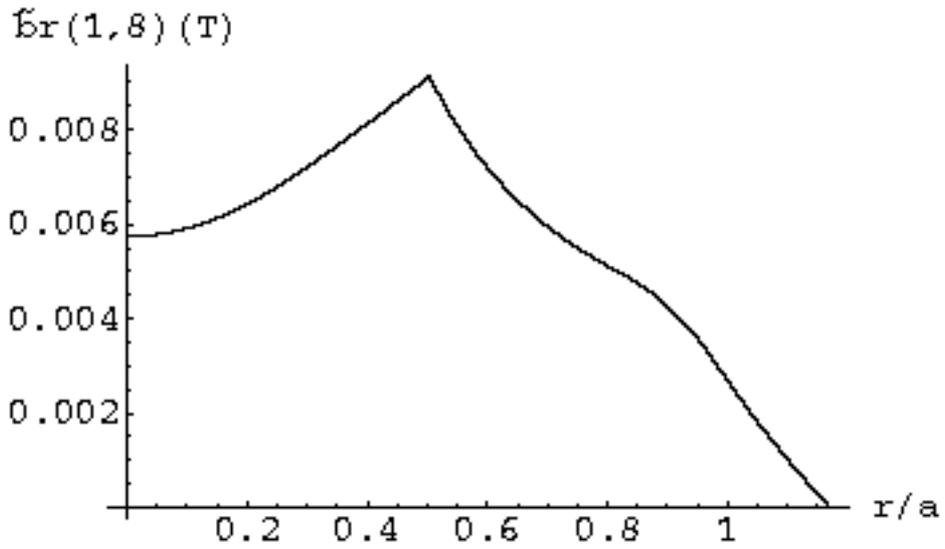

Fig. 10: The first-order radial field amplitude profile for the same shot of fig.9

Note the discontinuity of the first derivative across the mode rational surface. The radially integrated poloidal and toroidal component of the current sheet are therefore $\delta J_{\theta 1}^{1,8}$ = *7165A/m*, $\delta J_{\phi 1}^{1,8}$=*7781A/m*

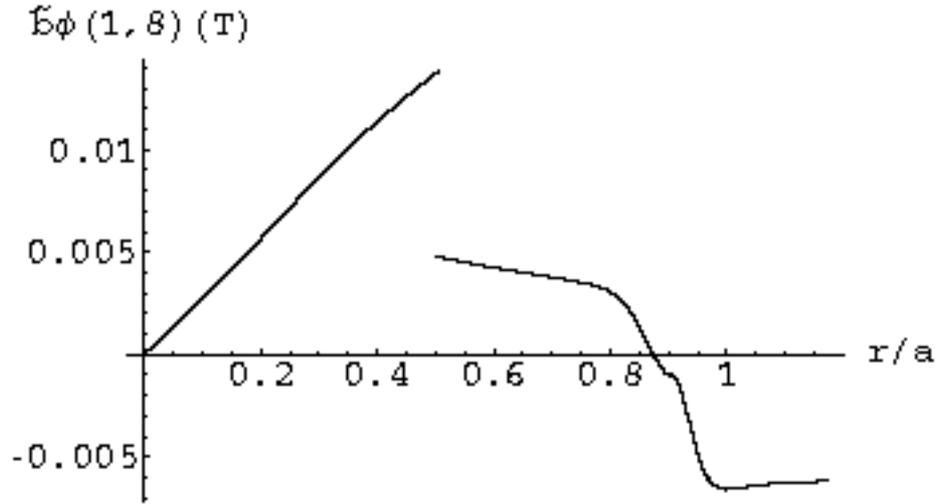

Fig. 11: The toroidal field first-order amplitude profile for the mode (1,8). The jump at the rational surface is a consequence of the presence of the $d\psi/dx$ discontinuity there (i.e., a current sheet flows at the rational surface).

For the same shot and time we consider the mode $m=1$, $n=14$. In this case the amplitude is lower: $b_\phi^{1,14}(1.17) = -1.75 \, 10^{-3} \, T$. The corresponding profiles are shown in figures 12, 13.

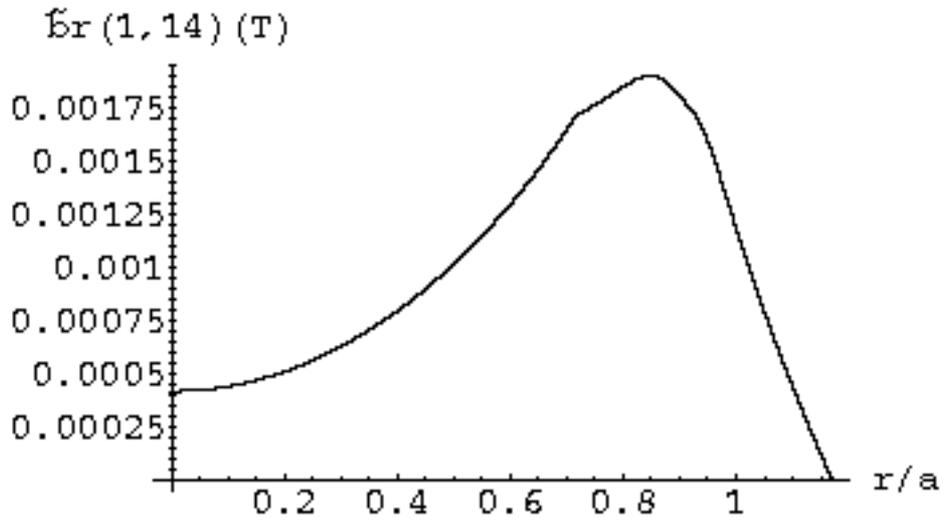

Fig. 12. Radial perturbation for the m=1, n=14 mode.

We find $\delta J_\theta^{1,14} = 391 \, A/m$, $\delta J_\phi^{1,14} = 169 \, A/m$.

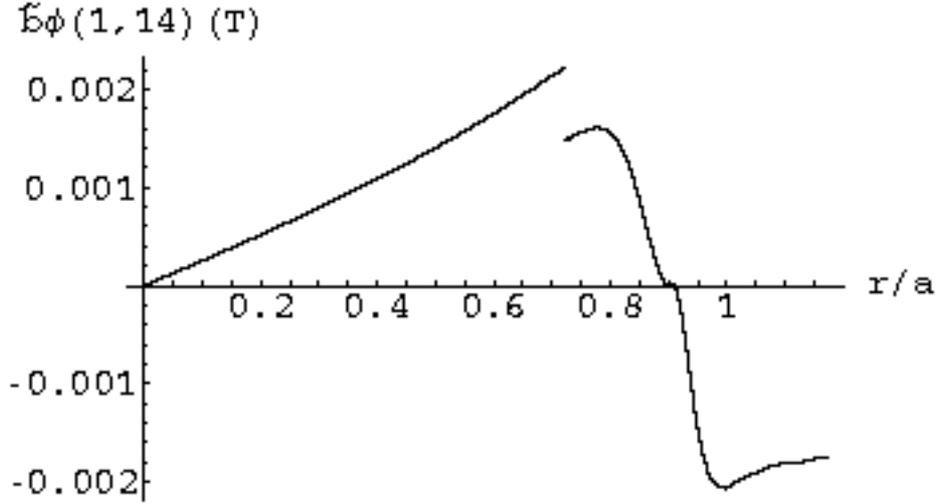

Fig. 13. Toroidal perturbation for the m=1, n=14 mode.

**6. Application of the model: particle and heat transport following stochastic line fields**

The particle and heat radial transport is a problem of fundamental importance in the context of performances of magnetically confined plasmas. It is well known that transport caused by Coulomb collisions in curved magnetic fields (neoclassical theory) only marginally accounts for transport in present tokamaks and is completely off the mark in present RFPs. This is ascribed to the destruction of magnetic surfaces due to MHD instabilities. Each mode, at the location of its resonance, leads to creation of magnetic islands and stochastic regions. The existence in RFPs of several modes, as we have just shown, leads to overlapping of the islands and hence to global stochasticity over a large fraction of the volume. Any given particle performs a ballistic motion along a field line but the overall displacement, due to the wandering nature of lines themselves, can be less than linearly increasing with time. Rechester and Rosenbluth [23] showed that, in a homogeneous magnetic field, the resulting motion along the radial direction is diffusive, with a diffusion coefficient

6.1) $\quad D \approx v_{th} \tilde{b}^2$

($v_{th}$ thermal velocity of the particle and $\tilde{b} = b_1 / B_0$ normalized amplitude of the symmetry-breaking magnetic perturbation). A precise comparison with empirical data is difficult within the naive formulation (6.1), because of the neglect of many realistic details. A full account is given in the paper by Maluckov *et al* [24], although they were presented already in [25], where an approximated version of our calculations was introduced. Briefly, Eq. (6.1) "forgets":

(I) Coulomb collisions. The particle motion is ballistic along the field line only until it is abruptly scattered on a different line by a collision with another charged particle. This effect is important in low-temperature highly collisional plasmas.

(II) Inhomogeneity of the magnetic field. This arises since fusionistic magnetic fields need a shear and, unavoidably, because of the finite size of the plasma. Authors [24,25] showed that the existence of boundaries has a dramatic qualitative effect on the long-time transport, altering its nature from diffusive to sub-diffusive (i.e., the mean squared displacement increases less than linearly with time): a field line can freely displace from its initial position only until reaches the proximity of the confining walls. After that, its motion is strongly constrained.

These limitations can be overridden only under some plasma regimes. Another obvious reason precluding the simple use of (6.1) is that the normalized perturbation amplitude $\tilde{b}$ should be known everywhere inside the plasma, and this was not possible: ordinarily, magnetic probes are used to measure fluctuations amplitudes only at the plasma edge. Our model is well suited for overcoming this difficulty: it provides the "true" perturbations all over the plasma, once just their boundary values are measured.

Once the total field **B** is known from the previous sections, the trajectory of a field line is recovered by solving the differential equations

6.2) $\dfrac{dx_i}{dl} = \dfrac{B_i}{B}$ or, equivalently $\dfrac{dx_i}{dt} = v\dfrac{B_i}{B}$ ($i = 1,...,3$)

where $dl$ is an infinitesimal path length, **x** the instantaneous position, **B(x)** the local total magnetic field (function of the position **x**). In the zero-mass limit, a particle follows exactly the field line. Its velocity is $v = dl/dt$. The zero-mass limit should be a rather good approximations for electrons, whose Larmor radius around the field line is much smaller than the typical scale of variation of **B**. Drift motion is neglected within this approximation, but it should not be a concern, since we are interested in the diffusive part of the motion.

The r.h.s. of (6.2) are implicit functions of **x**. Furthermore, field lines are likely to be chaotic, i.e., sensitive to any perturbation, either physical or numerical. For these reasons Eqns. (6.2) must be integrated using an accurate solver: we used a predictor-corrector scheme with a order-1 Adams-Bashfort algorithm for the predictor stage, and order-0 Adams-Moulton for the corrector stage.

A large number $N$ of trajectories are followed, starting from the same radial position and with random poloidal and toroidal positions. The following statistical quantities are recorded:

$$6.3) \quad \begin{cases} <\delta r(t)> = \dfrac{\sum_{i=1,N}(r_i(t)-r_i(0))}{N} \\ <\delta r^2(t)> = \dfrac{\sum_{i=1,N}(r_i(t)-r_i(0))^2}{N} - <\delta r(t)>^2 \end{cases}$$

In our investigation, we are interested to those regions where plasma is poorly collisional, hence we have neglected Coulomb collisions. They can perhaps be of importance for an accurate modelization of the outer, colder part of the plasma.

An ordinary diffusive motion would obey

$$6.4) \quad <\delta r^2> = 2Dt.$$

The diffusion coefficient $D$ is a measure of the global stochasticity of the plasma, hence -rigorously- it should be a simple number, constant over spatial coordinates. As we have said above, however, finite-size effects impose a crossover from the diffusive behaviour (6.4) for small times, to a sub-diffusive one ($<\delta r^2> = 2Dt^\alpha, \alpha<1$) for larger times. Sub-diffusive motion can be accommodated into generalized fluid continuity equations through the use of fractional derivatives. This conflicts, however, with the common practice of describing particle transport within the fairly intuitive "diffusion+convection" scheme. We resorted, therefore, to evaluate an effective diffusion coefficient by stopping the integration of (6.2) to times where Eq. (6.4) holds with a constant (i.e., time-independent) $D \equiv D_{\text{eff}}$. In practical terms, we fitted $<\delta r^2>$ of (6.4) with a straight line, limiting to regions where no sign of saturation was still visible. It is clear that, in this way, we are preventing the system from sampling the entire space. Hence, $D_{\text{eff}}$ comes from a local average around the starting position and is, therefore, function of the radius.

Prior to entering into quantitative calculations, an important comment is needed. Harvey *et al.* [26] showed that both particle and heat transport, in stochastic magnetic fields, are characterized by the same coefficient (6.1). However, because of the unavoidable insurgence of ambipolar electric fields, the effective particle transport is not determined by electrons, rather by main ions (provided that impurity content is negligible). That is, the effective (i.e., measured) particle diffusion coefficient is given by (6.1) with thermal velocity corresponding to the ionic one. The same constraint does not apply to thermal transport, for which electron velocity must instead be used. Within the above context, the terms "particle transport" and "heat transport" can be used interchangeably, through the simple scaling relation $D_{heat}/D_{part} = \sqrt{m_p/m_e} \approx 43$. It is known that in RFX this ratio does not rigorously hold, i.e., either heat or particle transport does not follow exactly Rechester-Rosembluth's predictions [28]; stating which one between them is difficult because of the large empirical errors. Indeed, any discussion on experimental error bars would be a tough one. Again, there are at least two main reasons for the large uncertainties involved in this kind of measurements: (I) first of all, no density local measurements are possible, but only inversion of line-of-sight measurements, on a relatively small number of chords; (II) transport coefficients are calculated as a flux-over-gradient ratio, e.g, $\chi = \Gamma_{heat}/\nabla T$ (we have used here the more customary symbol χ instead of $D_{heat}$). Temperature and particle profiles are very flat over a large fraction of the minor radius, in RFX, i.e., the denominator is almost zero. Hence, any uncertainty on it -to a large extent induced by the former point I- causes large errors on transport coefficients (for an estimate of heat transport coefficients and associated error bars, see [12, 27]).

### 7. Transport coefficients for some sample cases

For transport studies, we will consider the shot 11030, that features both standard confinement as well as a temporal window of good PPCD-induced improved confinement, and has been already used for PPCD studies [28]. PPCD (for Pulsed Poloidal Current Drive) is a well-established technique for improving RFPs' plasma parameters, although only transiently. It can induce an increase of electron temperature of order 30%-50% in the best cases. The physical mechanisms underlying this enhancement are not completely understood. The prevailing opinion is that PPCD replaces part of the spontaneous dynamo action of the plasma, hence allowing for a decrease of the magnetic fluctuations, a reduced overlapping of magnetic islands and, eventually, a dimished stochastic transport.

Figure 14 features the progressive destruction of magnetic surfaces for pulse 11029. From top to bottom, we have a poloidal cross section of the magnetic surfaces with: (I) no radial magnetic perturbation (i.e., only the **B**$_0$ field); (II) a single-mode perturbation (in this case, we have retained the *m*=1, *n* = 8 mode); (III) the full (*m* = 1) mode spectrum. All plots are referred at *t* = 30 ms (i.e., prior to the start of PPCD).

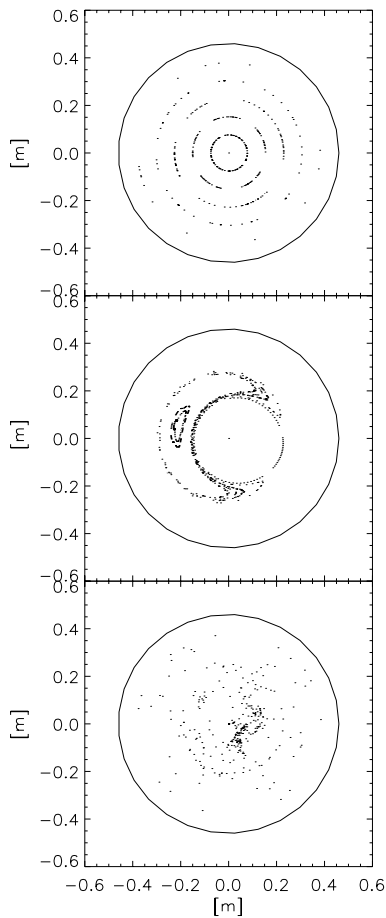

Fig. 14: Poincaré plots for pulse 11029. Upper panel, no perturbations; middle panel, just one mode; lower panel, full perturbation.

Figure 15 shows, instead, an example of $<dr^2>$ from Eq. (6.3). We can clearly distinguish two regions: one of linear increase (diffusive region) and another of saturated -or even reduced- displacement (subdiffusive region). A further region of superdiffusive motion ($<dr^2> \propto t^2$) can perhaps be identified at the beginning of the simulations, although it is more apparent in other runs. The diffusive region is the part of interest. In the course of our numerical investigations, we (re)discovered some unwanted features: the diffusive region decreases when one samples zones close to the center or to the edge. This is sensible, since we expect the boundaries to have a more marked effect as one approaches them. Indeed, they were already present in the paper [25]. However, the practical result is that the diffusive region quickly shrinks to an extent such that fitting becomes impossible. For this reason, we must limit our investigation to the middle region (say, $1/5 < r/a < 2/3$). This is not too severe a limitation. On the one hand, we have not reasons to expect drastical changes when going towards the core. On the other hand, the core region can be only badly diagnosed, hence any eventual numerical estimate should match empirical evaluations subject to fairly large error bars. On the other hand, the edge region, where data from the experiment are relatively better, is the ground of several other physical processes not taken into account within the present approach (high collisionality, electric field gradients, …) which dramatically affect transport. Above all, the edge region features the reversal surface where the $m = 0$ modes resonate. We have not taken them into consideration because: I) there is a lot of them, thus making calculations burdensome, and II) their amplitude is rather small, if compared to $m = 1$ modes, hence their effect -far from the reversal surface- is likely to be negligible.

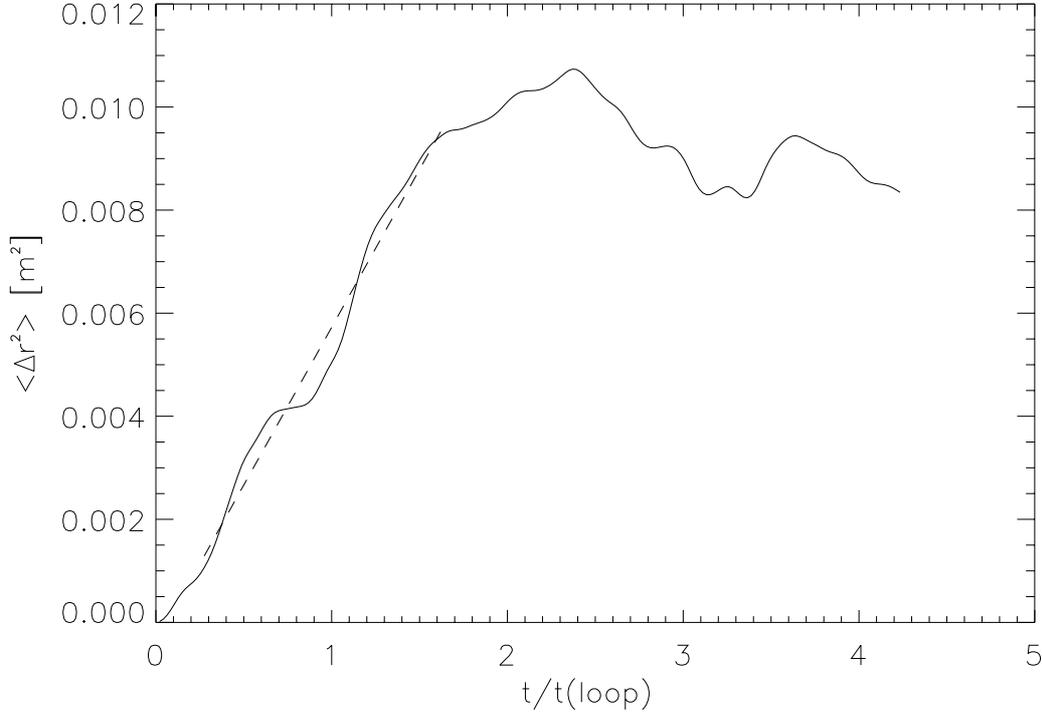

Fig. 15: mean square radial displacement versus time.

We have plotted in Figure 16 the resulting radial profile of effective diffusion coefficient (6.4) (linear fit to $<dr^2>$ data), together with the inferred one from interferometry data, using an interpretative code [29], and from a heat transport analysis [12, 28]. This, for two instants of time, one prior and one during PPCD. We plot in Fig. 17 the corresponding first-order amplitudes $b_r$ for all of the ($m = 1$) modes involved. The prevailing of one single mode over all the others is dubbed Quasi-Single-Helicity (QSH). Its theoretical limit - the Single Helicity state- involves the complete disappearance of all modes but one. This limit has been theoretically predicted as a potential RFP state fulfilling Cowling's theorem but without the need of turbulence for the generation of dynamo field, hence a very appreciable situation [30,31]. In our case we are in a situation akin to QSH states: one mode (n = 7) increases in the central region of the plasma at the expenses of the secondary (n ≥ 9) modes.

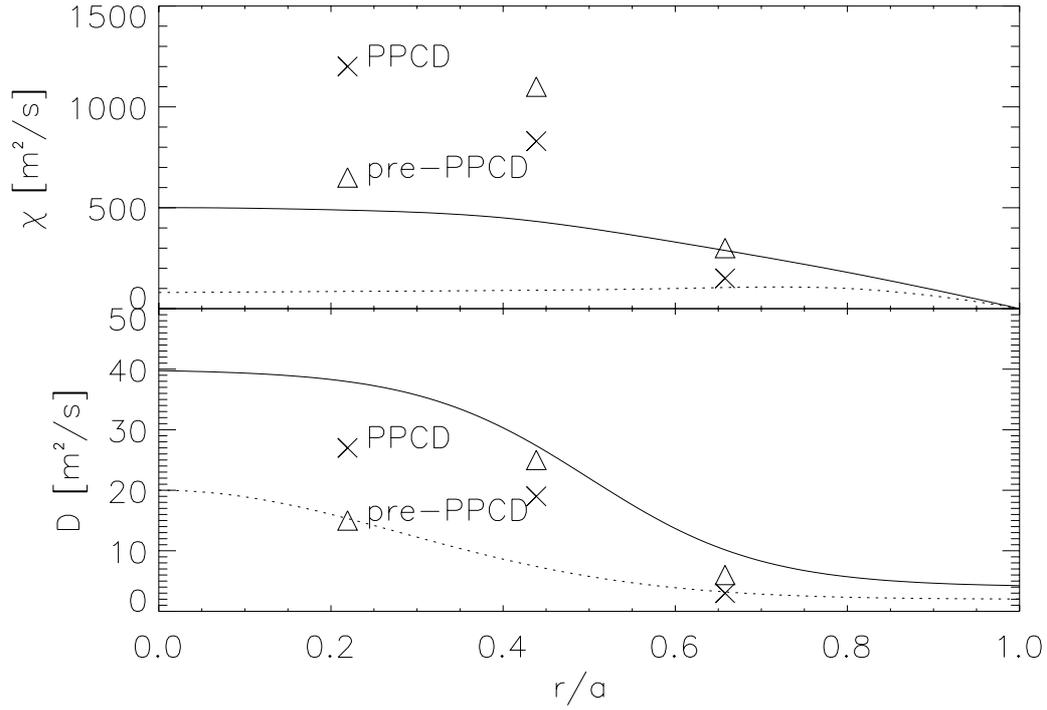

Fig. 16: upper plot, heat transport coefficient; lower plot, particle transport coefficient, from experiment and for pulse 11030 Solid curve, t = 30 ms (pre-PPCD); dotted curve, t = 35 ms (PPCD). Triangles and crosses, results from the model.

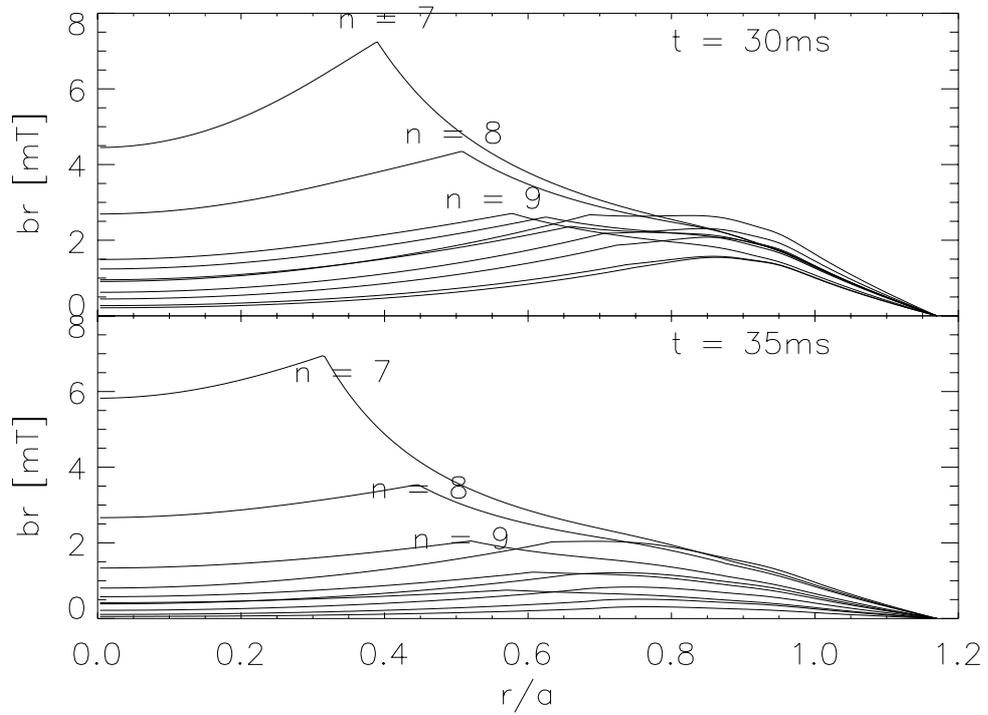

Fig. 17: First-order amplitude of the radial component of perturbations for shot 11030 at two time points.

The disappearance of the small outermost modes during PPCD reflects on a decrease of $D$, $\chi$ in the outer region. By contrast, in the core region, transport during PPCD is even enhanced, since the two main modes (n=7,8) increase there.

Indeed, it is clear any agreement between theory and experiment can only be qualitative. There are a few points that deserve to be stressed: I) experimentally, it is hardly possible to identify a non-monotonic $D$, $\chi$ profile, because of the lack of mesh resolution. This means that the decrease of these coefficients when going towards the origin -in the pre-PPCD case- cannot be ruled out, although it is not supported by experimental findings. II) the decrease of the transport coefficients during PPCD is not as dramatic as suggested by experiment on the whole radius, instead it features more a transport barrier (this will appear more clearly below).

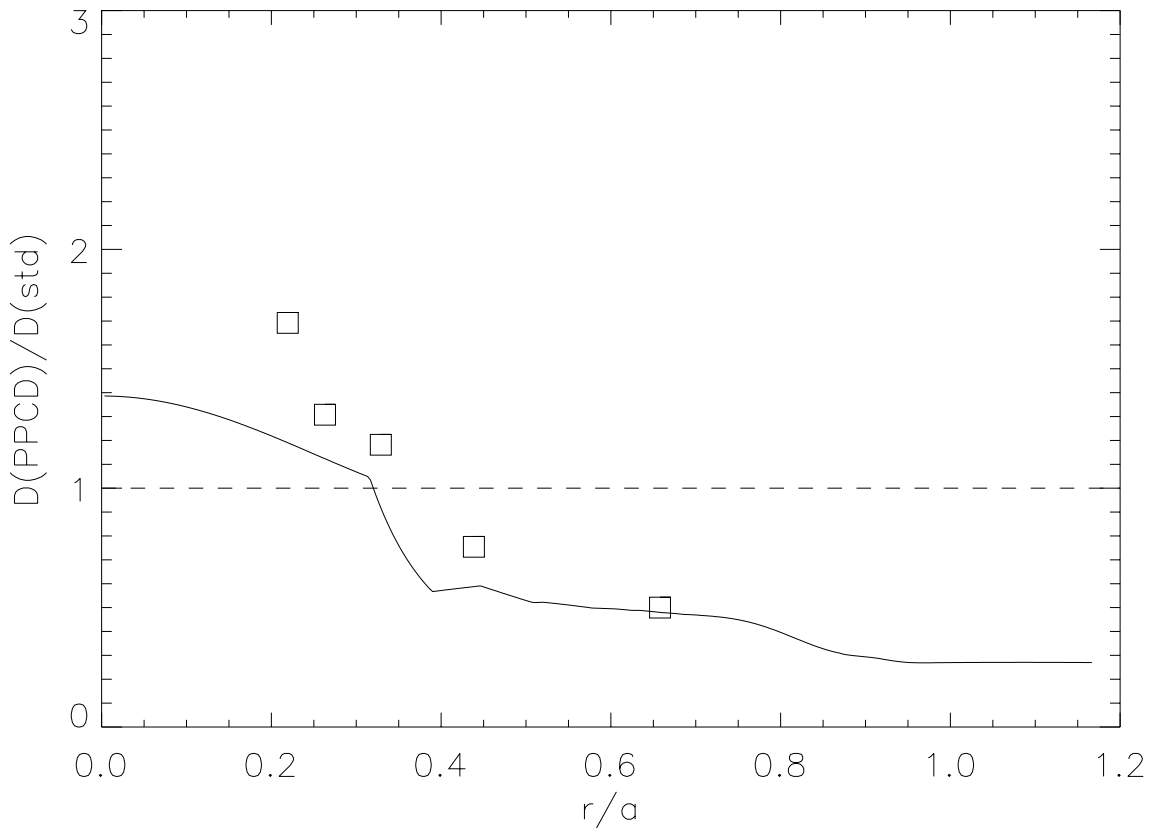

Fig.18: squares, ratio between numerically computed transport coefficient: D(PPCD)/D(pre-PPCD); solid line, ratio between normalized radial field perturbation amplitudes $|b_{rl}^2|$, evaluated at the same times. The dashed line represent the condition for which PPCD has no effect on transport.

In order to give a further intuitive picture, together with a test about the correctness of our results, we plot in Fig. 18 our computed ratio between transport coefficients during and before PPCD (squares). According to equations above, this ratio should be weakly dependent on geometric and/or plasma factors, but the normalized perturbations amplitudes: roughly, $\frac{D(t_1)}{D(t_2)} \approx \frac{|b_r^2(t_1)|}{|b_r^2(t_2)|}$. The solid line in Fig. 18 is the r.h.s. of this equation. We see that our expectation is rather well fulfilled but for the points $r/a \leq 0.2$. This is due to the fact that, in this region, our procedure was not able to give a reliable estimate for $D$ prior to PPCD: the $<dr^2>$ versus $t$ curve was never a true straight line but there was almost only a smooth crossover to saturation. Therefore, unavoidably, part of the fitting collected points that pertained to the subdiffusive region. The resulting slope was below its true value. The same did not appear so clearly for the PPCD-case, hence the effective $D$ was estimated better.

From the point of view of physics, it is remarkable to notice that the improved confinement region (where the ratio is less than one) is not spread all over the radius but only over a region of it ($r/a > 0.3$). This is at a difference with the most common picture of PPCD, which foresees a global improvement of transport.

### 8. Conclusions

The equilibrium in a finite-$\beta$ RFP plasma in the presence of saturated-amplitude tearing modes has been investigated analytically and numerically. The singularities of the force balance equation at the modes rational surfaces have been dealt with by a proper regularization of the zeroth-order cylindrically symmetric profiles: the gradients of the pressure and parallel current density ($\sigma$) are forced to be equal to zero there. This rule has been combined with some considerations about general bounds plasmas are constrained to, to guess equilibrium profiles which match the global discharge parameters $(F, \Theta, \beta_p)$. The combined effects of a large number of modes makes the $\sigma$ and pressure profiles flat over a large portion of the plasma, with a steep gradient in the edge region, although non-monotonic $\sigma$ profiles, with a maximum at the reversal surface, are also allowed in principle. A corollary of our regularization rule is that to obtain steeper profiles an RFP needs to reduce drastically the dynamo modes amplitude. This is not surprising, if we think that all the improved regime of operations in RFX ($\alpha$-mode [11], PPCD [12], "single-helicity" states [30]) are characterized by a reduced mode amplitude.

At present, in RFX we do not have a shot-by shot measurement of σ and pressure profile. Anyway, the first experimental analysis of σ(r) [10] is largely compatible with our model. The same is true for the pressure, even if only the electronic component is measured [13]. When some of the modes are suppressed, the profiles both in our model and in the experiment become steeper.

The model can be regarded as a method to obtain the final stage of relaxation process produced by the tearing modes. This equilibrium model allows the use of a relatively simple equation to get the various perturbations profiles. In this way a complete magnetic reconstruction is obtained.

Some of the reasonings which have led us to the results shown in previous sections, are forcefully semi-quantitative, therefore some room to arbitrariness is left. In particular, we think that finding further constraints allowing to better define "shape" functions $w, u$ (Eq. 3.12), would be a possible improvement. Furthermore, the present version of our model allows only an "on/off" picture of the flattening effect of the resonant modes over the equilibrium. In a more sophisticated model a correlation between the mode amplitude and the width of the induced flattening should be self-consistently taken into account.

As it is, the model already lends itself to applications. The calculation of transport coefficients in a RFP is, indeed, an important issue. The results we have obtained in sections 6,7 must be considered as preliminary, since there is a large uncertainty both about them, as well as about the empirical data they have to be compared to. Notwithstanding this, they appear encouraging since they are not grossly away from experiment. Also, if verified by further analysis, they show a picture of transport reduction by PPCD which is slightly different from the standard one, i.e., instead of an overall decrease of transport along the whole radius, one gets a rather localized one, corresponding to the outer region where smaller modes collapse, with some affinities to transport barriers in tokamaks.

## Appendix

This appendix attempts a different approach to the equilibrium in the presence of resonant perturbations, and confirms the results obtained in Section 2.

Let's assume the existence of an island centred on the *(m,n)* rational surface. To the lowest order the standard representation for the helical magnetic flux in the island rest frame is [32]:

A1) $\Psi(r,\zeta) = \Psi_0(r) + \Psi_1(r)\cos\zeta; \quad \Psi_0(r) = -\int_{r_s^{m,n}}^{r} F^{m,n}(\rho)d\rho; \quad \Psi_1(r) = \tilde{\psi}^{m,n}(r);$

where $\zeta = m\theta - n\phi$ is the helical angle. The island width is $W = 4\left(\Psi_1/(dF^{m,n}/dr)\right)^{1/2}\Big|_{r_s^{m,n}}$. This representation holds in the vicinity of the rational surface. Let's define for each quantity the

cylindrical component $\overline{C} = \frac{1}{2\pi}\oint d\zeta\, C$ and the fluctuating component $\tilde{C} = C - \overline{C}$. From $\mathbf{B}\cdot\nabla p = 0$ we get $p=p(\Psi)$ and therefore

A2) $\quad \dfrac{\partial p}{\partial r}\dfrac{\partial \Psi}{\partial \zeta} - \dfrac{\partial p}{\partial \zeta}\dfrac{\partial \Psi}{\partial r} = 0$

Inserting (A1), the latter condition becomes:

A3) $\quad \dfrac{\partial \tilde{p}}{\partial r}\Psi_1 \sin\zeta + \dfrac{\partial \tilde{p}}{\partial \zeta}\left(\dfrac{\partial \Psi_0}{\partial r} + \dfrac{\partial \Psi_1}{\partial r}\cos\zeta\right) = 0$

At the rational surface $\left.\dfrac{\partial \Psi_0}{\partial r}\right|_{r_s^{m,n}} = 0$. Moreover $\left.\dfrac{\partial \Psi_1}{\partial r}\right|_{r_s^{m,n}} = 0$, because for a tearing mode the perturbed flux is even across the rational surface [33]. Therefore $\left.\dfrac{\partial p}{\partial r}\right|_{r_s^{m,n}} = \left.\dfrac{\partial \overline{p}}{\partial r}\right|_{r_s^{m,n}} = \left.\dfrac{\partial \tilde{p}}{\partial r}\right|_{r_s^{m,n}} = 0$ so the pressure derivative is zero at the rational surface, both for the cylindrical and the fluctuating part. As done in Section 2 we impose that the cylindrical component of the pressure is a monotonic function nearby the rational surface: $\left.\dfrac{\partial^2 \overline{p}}{\partial r^2}\right|_{r_s^{m,n}} = 0$. Taking the radial derivative of (A3) at the rational surface, the latter condition brings to

A4) $\quad \left.\left(\dfrac{\partial^2 \tilde{p}}{\partial r^2}\Psi_1 \sin\zeta + \dfrac{\partial \tilde{p}}{\partial \zeta}\dfrac{\partial^2 \Psi_1}{\partial r^2}\cos\zeta - \dfrac{\partial \tilde{p}}{\partial \zeta}\dfrac{dF^{m,n}}{dr}\right)\right|_{r_s^{m,n}} = 0$

Note that the first and second terms are higher order with respect to the third. The only way to satisfy (A4) is therefore is to impose $\left.\partial \tilde{p}/\partial \zeta\right|_{r_s^{m,n}} = 0$, $\left.\partial^2 \tilde{p}/\partial r^2\right|_{r_s^{m,n}} = 0$ at the rational surface. Since

A5) $\quad \dfrac{\partial p}{\partial \zeta} = -\dfrac{dp}{d\Psi}\Psi_1 \sin\zeta$

this is possible only if $dp/d\Psi=0$ for $r=r_s^{m,n}$, $0<\zeta<\pi$. From (A1) this means $dp/d\Psi=0$ in the interval $-\Psi_1(r_s^{m,n}) < \Psi < \Psi_1(r_s^{m,n})$, which covers the entire island region. Therefore in this region we get a constant pressure, and the equilibrium becomes force free:

A6) $\mathbf{J} \times \mathbf{B} = \nabla p = 0 \quad \Rightarrow \quad \mu_0 \mathbf{J} = \mu \mathbf{B}$

Th function $\mu(r,\zeta)$ generalizes the cylindrical parameter $\sigma(r)$ defined in Section 2. From $\nabla \cdot \mathbf{J} = 0$ we get $\mathbf{B} \cdot \nabla \mu = 0$, so $\mu=\mu(\Psi)$. To this function we can apply the same conditions (A2, A3) and get $\left.\frac{\partial \mu}{\partial r}\right|_{r_s^{m,n}} = \left.\frac{\partial \overline{\mu}}{\partial r}\right|_{r_s^{m,n}} = \left.\frac{\partial \tilde{\mu}}{\partial r}\right|_{r_s^{m,n}} = 0$. The regularization rule (2.38) is therefore recovered.

### References

[1] S. Ortolani and D.D. Schnack, *Magnetohydrodynamics of Plasma Relaxation*, (World Scientific, Singapore, 1993)

[2] J.M. Finn, R. Nebel, C. Bathke, Physics of Fluids B **4** (1992) 1262

[3] R. Fitzpatrick and P. Zanca, Physics of Plasmas **9** (2002) 2707

[4] H.P. Furth, J. Killen, and M.N. Rosenbluth, Physics of Fluids **6** (1963) 459

[5] D. Merlin, *et al*., Nuclear Fusion **29** (1989) 1153

[6] L.S. Solov'ev and V.D. Shafranov, Reviews of Plasma Physics, Vol. 5 (1970), ch. 1

[7] T.C Hender, D.C. Robinson, Nuclear Fusion **21** (1981) 755; R. Fitzpatrick, Physics of Plasmas **2** (1995) 825; C. Ren, *et al*., Physics of Plasmas **5** (1998) 450

[8] J. B. Taylor, Reviews of Modern Physics **58** (1986) 741

[9] G. Rostagni, Fusion Engineering and Design **25** (1995) 301

[10] M. Bagatin, *et al*., Review of Scientific Instruments **72** (2001) 426

[11] S. Martini, *et al*., Plasma Physics and Controlled Fusion **41** (1999) A315

[12] R. Bartiromo, *et al*., Physical Review Letters **82** (1999) 1462

[13] T. Bolzonella, *et al*., Physical Review Letters **87** (2001) 195001

[14] D.L. Brower, *et al*., Physical Review Letters **88** (2002) 185005

[15] D.Brotherton-Ratcliffe, C.G.Gimblett, I.H.Hutchinson, Plasma Physics and Controlled Fusion **29** (1987) 161

[16] P. Zanca and F. Sattin, Plasma Physics and Controlled Fusion **45** (2003) 1

[17] D.C. Robinson, Nuclear Fusion **18** (1978) 939

[18] R. Fitzpatrick, *et al*., Nuclear Fusion **33** (1993) 1533

[19] J. P. Freidberg, *Ideal Magnetohydrodynamics*, (Plenum Press, New York and London, 1987)

[20] R. Fitzpatrick, Physics of Plasmas **6** (1999) 1168

[21] G. Marchiori, private communication (2002)

[22] P. Zanca, *et al*., Physics of Plasmas **8** (2001) 516

[23] A.B. Rechester and M.N. Rosenbluth, Physical Review Letters **40** (1978) 38

[24] A. Maluckov, *et al.,* Physica A **322** (2003) 13

[25] F. D'Angelo and R. Paccagnella, Physics of Plasmas **3** (1996) 2353

[26] R.W. Harvey, M.G. McCoy, J.Y. Hsu, and A.A. Mirin, Physical Review Letters **47** (1981) 102

[27] A. Intravaia, *et al*., Physical Review Letters **83** (1999) 5499

[28] M.E. Puiatti, *et al*., Nuclear Fusion **43** (2003) 1057

[29] D. Gregoratto, *et al.,* Nuclear Fusion **38** (1996) 1199

[30] D.F. Escande, *et al.*, Plasma Physics and Controlled Fusion **42** (2000) B243

[31] P. Martin, *et al*., Physics of Plasmas **7** (2000) 1984

[32] P. H. Rutherford, Physics of Fluids **16** (1973) 1903

[33] R. Fitzpatrick, Phys. Plasmas **1** (1994) 3308